\documentclass[%
reprint,
 amsmath,amssymb,
 aps,
 prl,
]{revtex4-1}

\usepackage{graphicx}
\usepackage{dcolumn}
\usepackage{bm}

\begin{document}

\preprint{APS/123-QED}

\title{Shiba Bound States across the mobility edge in doped InAs nanowires}

\author{Alexandre Assouline}
\affiliation{LPEM, ESPCI Paris, PSL Research University; CNRS; Sorbonne Universit\'es, UPMC University of Paris 6\\
10 rue Vauquelin, F-75005 Paris, France}

\author{Cheryl Feuillet-Palma}
\affiliation{LPEM, ESPCI Paris, PSL Research University; CNRS; Sorbonne Universit\'es, UPMC University of Paris 6\\
10 rue Vauquelin, F-75005 Paris, France}

\author{Alexandre Zimmers}
\affiliation{LPEM, ESPCI Paris, PSL Research University; CNRS; Sorbonne Universit\'es, UPMC University of Paris 6\\
10 rue Vauquelin, F-75005 Paris, France}

\author{Marco Aprili}
\affiliation{Laboratoire de Physique des Solides, CNRS, Univ. Paris-Sud, University Paris-Saclay, 91405 Orsay Cedex, France}

\author{Jean-Christophe Harmand}
\affiliation{Centre de Nanosciences et de Nanotechnologies, CNRS, Univ. Paris-Sud, Universit\'es Paris-Saclay, C2N – Marcoussis, 91460 Marcoussis, France}

\author{Herv\'e Aubin}
\email{Herve.Aubin@espci.fr} 
\affiliation{LPEM, ESPCI Paris, PSL Research University; CNRS; Sorbonne Universit\'es, UPMC University of Paris 6\\
10 rue Vauquelin, F-75005 Paris, France}

\date{\today}

\begin{abstract}
We present a study of Andreev Quantum Dots (QDots) fabricated with small-diameter (30 nm) Si-doped InAs nanowires where the Fermi level can be tuned across a mobility edge separating localized states from delocalized states. The transition to the insulating phase is identified by a drop in the amplitude and width of the excited levels and is found to have remarkable consequences on the spectrum of superconducting SubGap Resonances (SGRs). While at deeply localized levels, only quasiparticles co-tunneling is observed, for slightly delocalized levels, Shiba bound states form and a parity changing quantum phase transition is identified by a crossing of the bound states at zero energy. Finally, in the metallic regime, single Andreev resonances are observed.
\end{abstract}

\pacs{Valid PACS appear here}
\maketitle

In superconductor-QDot-superconductor structures or at impurities in bulk superconductors, Bogoliubov type bound states can form at energy smaller than the superconducting gap energy $\Delta$, leading to SGRs.

The SGR formation depends on the different energy scales: $\Delta$, the coupling $\Gamma_{\rm S}$ of the QDot with the superconducting electrodes, the charging energy, U, and the energy, $\varepsilon_0$, of the QDot level relative to the Fermi energy of the superconducting electrodes.
Its phase diagram has been extensively studied theoretically\cite{Rozhkov1999-hd,Martin-Rodero2011-ji}. For large coupling $\Gamma_{\rm S}$, the SGRs result from the coherent superposition of multiple Andreev reflections\cite{Kulik1969-ho,Octavio1983-bj} and conductance peaks are expected at voltage values $\Delta$/ne\cite{Scheer1998-ga} where n is the number of Andreev reflections. For weak coupling $\Gamma_{\rm S}$, where the system is in the regime of Coulomb blockade, the ground state, singlet $|S\rangle$ or doublet $|D\rangle$, results from the competition between the Kondo screening and the superconducting pairing interaction. These two ground states are separated by a parity changing quantum phase transition, which can be identified by the crossing of the SGRs at zero-energy. Previous works have addressed this transition through measurements of Josephson supercurrents\cite{Van_Dam2006-vi,Cleuziou2006-he,Jorgensen2007-xc,Maurand2012-tb,Kanai2010-wc} and studies of the SGRs in S-QDot-S \cite{Eichler2007-lq,Sand-Jespersen2007-zp,Pillet2010-wt,Deacon2010-ub,Giazotto2011-ra,Pillet2013-lr,Kumar2014-mx} or N-QDot-S geometries\cite{Dirks2010-ll,Deacon2010-pq,Lee2012-ew,Chang2013-wq}. Recently, similar devices attracted intense interest with the observation of the zero-energy Majorana end states in proximitized nanowires\cite{Mourik2012-ij,Das2012-hm,Deng2012-mb}. 



The physics of odd parity QDot is related to the physics of Shiba states\cite{Yu1965-bp,Shiba1968-xa,Rusinov1969-fa,Soda1967-dt,Shiba1969-zs} forming at magnetic impurities in bulk superconductors, where tuning the magnetic exchange also leads to a parity changing quantum phase transition\cite{Sakurai1970-ye,Satori1992-uc,Simonin1995-rk,Salkola1997-ey,Balatsky2006-mp,Bauer2007-yn,Tsai2009-mk}, characterized by a crossing of the Shiba states at zero energy\cite{Franke2011-fw}.

So far, the impurity induced superconducting SGRs have been only observed by Scanning Tunneling Microscopy, at Co atoms deposited on Nb\cite{Yazdani1999-ww}, at magnetic impurities in 2H-NbSe$_2$\cite{Menard2015-yl} and magnetic MnPc molecules deposited on Pb\cite{Franke2011-fw}. However, tunneling spectroscopy of impurity states can also be done with microfabricated devices, such as nano-sized Schottky diodes~\cite{Calvet2002-lp}, nanosized field effect transistors~\cite{Sellier2006-ru,Calvet2011-ew,Calvet2007-fz, Calvet2007-jm} or nanosized memristors\cite{Mottaghizadeh2014-hn}. In this work, we present an observation of Shiba bound states forming in a dopant-induced impurity band. In this diffusive regime, it is expected that an Anderson-like Metal-Insulator Transition (MIT)\cite{Lee1985-pe} separates the metallic regime at high carrier concentration from an insulating regime at low carrier concentration, where the localized and delocalized states in the band structure are separated in energy by a mobility edge. We identified this mobility edge in the conductance spectrum measured as function of gate voltage and show  that the SGRs are sensitive to the metal-insulator transition through the effect of localization on the coupling parameter $\Gamma_{\rm S}$.

\begin{figure}[ht!]
	\begin{center}
		\includegraphics[width=7cm]{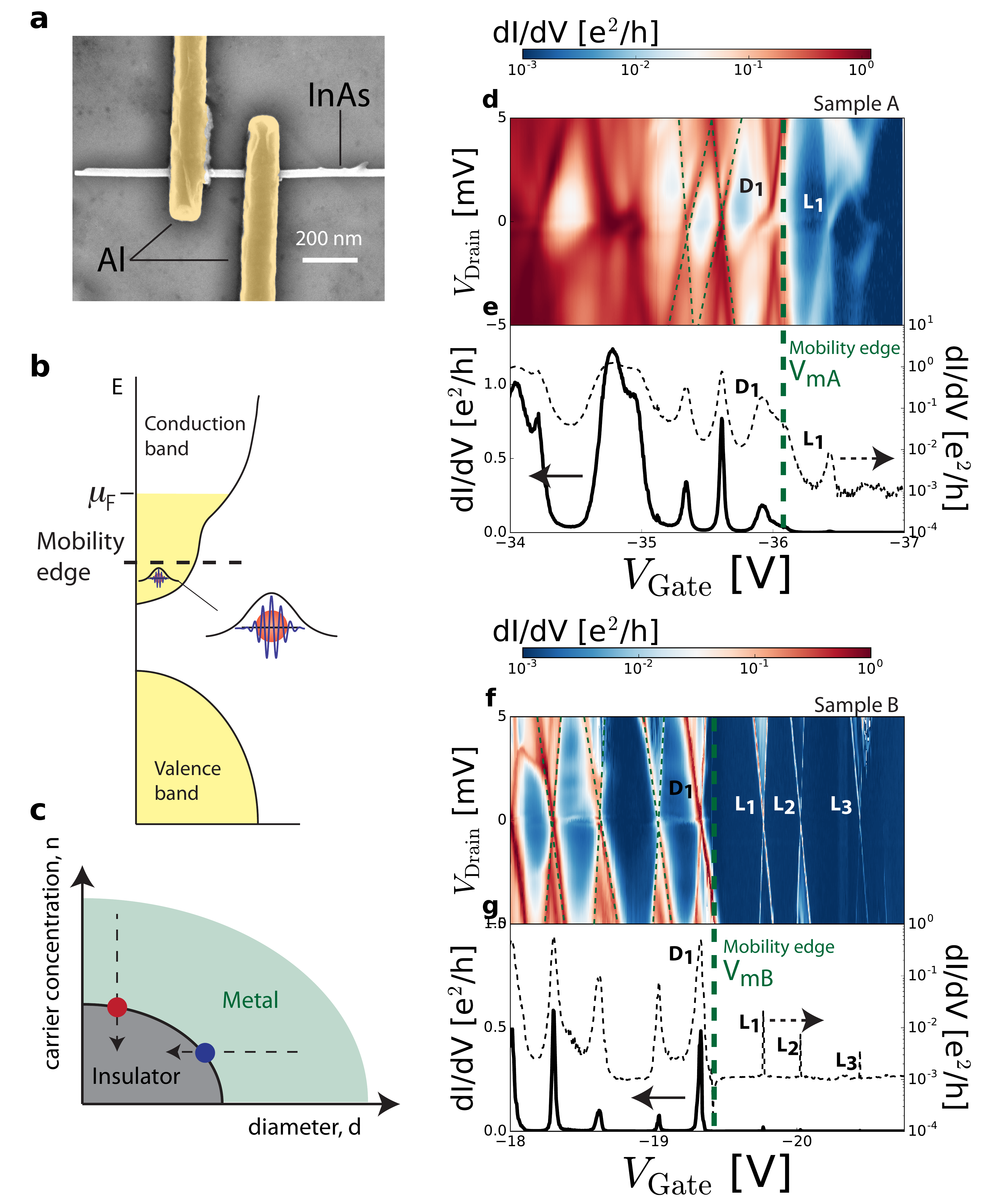}
		\caption{\label{Fig1} a) SEM image of a nanowire device. b) Sketch of the band structure showing the formation of an impurity band merging with the conduction band. c) Phase diagram displaying MIT as function of carrier density (red dot) and nanowire diameter d (blue dot). d,f) Conductance dI/dV maps for nanowires A and B, respectively, as function of drain bias $V_{\rm Drain}$ and back-gate voltage $V_{\rm Gate}$. Coulomb diamonds are highlighted with dash lines. These maps show the crossing of the mobility edge with the back-gate voltage, below which only a few excited levels remain visible, labelled L$_1$ to L$_3$. The last delocalized level for each nanowire is labelled D$_1$. e,g) Zero-bias conductance as a function of back gate voltage plotted on linear scale (continuous line) and log-scale (dash line).}
	\end{center}
\end{figure}

Epitaxially grown InAs nanowires of diameter approximately d=30 nm\cite{Glas2007-ch} were deposited on a degenerately doped Si substrate covered with a 300 nm thick thermal oxide. The nanowires are contacted with e-beam deposited Ti(5 nm)/Al (100 nm) electrodes with moderate in-situ ion beam cleaning of the nanowire surface before deposition. Fig.~\ref{Fig1}a shows an SEM image of one device. About 20 devices have been fabricated and measured with a standard lock-in method. The room temperature resistivity is below 10 m$\Omega$.cm, which is 20 times smaller than the value measured for undoped nanowires of similar diameter\cite{Scheffler2009-nm}. The two-wires and four-wires resistance are similar, indicating negligible contact resistance. Two devices, sample A and sample B, have been extensively measured in a dilution fridge with a base temperature of 30 mK. Upon cooling, the resistivity increases only up to  15 m$\Omega$.cm, which is 2$\times10^4$ smaller than measured on undoped nanowires, indicating metallic behavior as shown now.

\begin{figure}[ht!]
	\begin{center}
		\includegraphics[width=7cm]{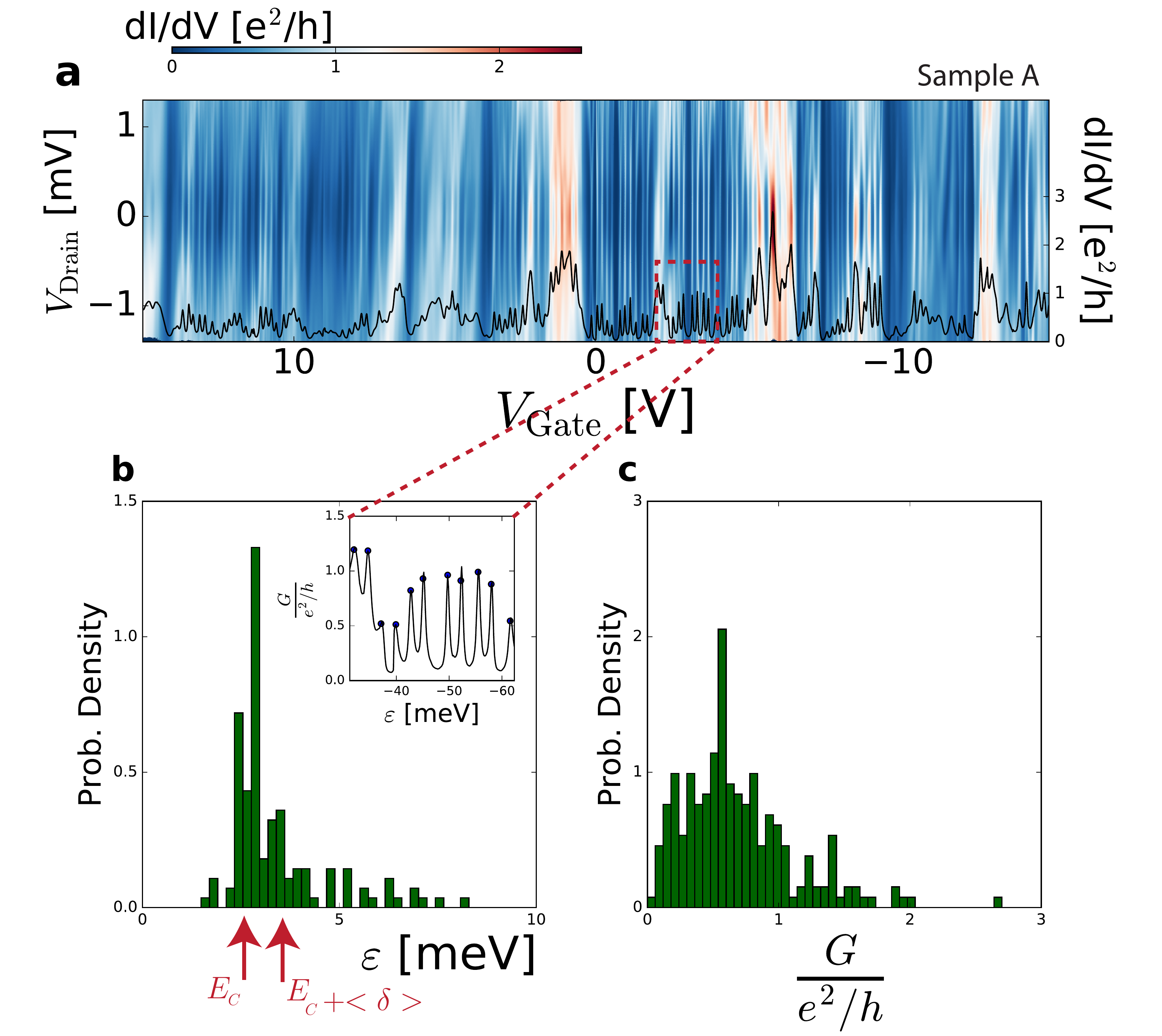}
		\caption{\label{Fig2} a) Conductance dI/dV maps for nanowire A as function of drain bias $V_{\rm Drain}$ and back-gate voltage $V_{\rm Gate}$. b) Histogram of level spacing obtained from the energy separation of the Coulomb peaks (inset). As indicated by the red arrows, the main peak of the histogram is located at the charging energy $E_{C}$. A second smaller peak is also expected at $E_{C}+\langle\delta\rangle$. Because of insufficient histogram resolution, a full comparison with the Wigner bi-modal distribution has not been attempted. c) Histogram of the Coulomb peak conductance maxima measured at zero drain bias.}
	\end{center}
\end{figure}

Fig.~\ref{Fig1}d and Fig.~\ref{Fig1}f show conductance maps dI/dV as function of drain and gate bias for sample A and sample B, respectively. They are measured with a large AC voltage ($\simeq 50~\mu$V) on a large drain voltage range, making the superconducting-like features barely visible. They show Coulomb blockade diamonds from which the lever arms $\alpha_{D,S,G}=C_{\rm D,S,G}/C_{\sum{}}$ and the corresponding capacitances $C_{\rm D,S,G}$ are obtained\cite{Ihn2004-gn}. The Coulomb energy extracted from the gate voltage separation between nodes $\Delta V_{\rm G}$ is given by $E_{\rm C}$=e$\alpha_{\rm G} \Delta V_{\rm G}\simeq 2.6~$meV for sample A, E$_{\rm C}$=4.4 meV for sample B.

Upon depleting the nanowire with a negative back-gate voltage, one can identify a threshold, $V_{\rm mA}$=-36 V, $V_{\rm mB}$=-19.4 V, for nanowires A and B, Fig.~\ref{Fig1}e and Fig.~\ref{Fig1}g, respectively, below which the excitation lines, labelled L$_1$ to L$_3$, are much narrower and their amplitude much weaker than the excitation lines above the threshold. For nanowire B, the amplitude of the Coulomb peak maxima $A_{\rm L_1}$=$A_{\rm L_2}$=0.12, $A_{\rm L_3}$=0.01, normalized to the amplitude of the last level D$_1$ above the threshold, decreases with the energy of the levels, indicating that the lower energy levels are more strongly localized. This striking evolution across the threshold indicates that it corresponds to the mobility edge where the states of energy above the mobility edge are delocalized and the states below are localized due to Anderson localization. For both samples, no other excitation levels are observed at gate voltages lower than the localized levels, indicating that the nanowires are indeed completely depleted at these most negative voltages, as shown on the zero bias conductance measured on a large gate voltage range shown Fig.~S1. In past works on phosphorus doped nanosized metal-Si-metal structures\cite{Sellier2006-ru}, dopant levels have also been identified as sharp conductance lines below the pinch-off voltage where the channel is fully depleted.

From these voltage thresholds, an estimation of the carrier concentration at near-zero gate bias is obtained from the relation n=C$_G$V$_{\rm mA(B)}$/(el$\pi d^2/4$), where l is the distance between the electrodes. Ones find $n_{\rm A}=1.7\times10^{18}$ cm$^{-3}$ for sample A (l=100 nm) and $n_{\rm B}=2.1\times10^{18}$ cm$^{-3}$ for sample B (l=200 nm). This carrier concentration is 10 times larger than in undoped nanowires\cite{Scheffler2009-nm}. Assuming that each Si atom provides one electron, each QDot contains $\sim 135$ dopants per 100 nm unit length. For this density, the average distance between dopants is $\langle r \rangle \simeq 8~$nm, which is smaller than the Bohr radius $a_0\simeq 30.2~$nm, calculated from the relation $a_0=4\pi\varepsilon_0\varepsilon\hbar^2/m^*e^2$ with $\varepsilon =~15.15$ and the effective mass $m^*=0.0265$. In other words, the carrier concentration is much larger than critical density $N_{\rm c}=5.6\times10^{14}$ cm$^{-3}$ obtained from the Mott criterion\cite{Mott1990-ga} $a_{\rm B}N_{c}^{1/3}\simeq$0.25 for metallicity. At this high doping level, an impurity band is expected to form and merge with the conduction band, as sketched Fig.~1b, Ref.~\cite{Schubert2015-ne}. As discussed by Mott\cite{Mott1990-ga}, the Ioffe-Regel criterion for metallicity $\lambda_F^{-1}\ell\simeq 1$ implies that a mobility edge separates the delocalized states ($\lambda_F<\ell$) at high energy from the Anderson localized states ($\lambda_F>\ell$) at low energy. Thus, a MIT is expected either as function of disorder or carrier concentration as sketched on the phase diagram Fig.~1c. In nanowires, the amount of disorder is controlled by the diameter, as shown in Ref.~\cite{Scheffler2009-nm} where the mobility of \emph{undoped} nanowires drops abruptly below a critical diameter about ~50 nm, which corresponds to a disorder driven MIT. In contrast, while our nanowires have a diameter smaller than this critical diameter, they are metallic because of their high carrier concentration. They are driven into the insulating regime upon reducing the carrier concentration with the gate voltage.

An indication that the excitation levels observed above the mobility edge are also part of the impurity band comes from the level spacings, which is expected to show large fluctuations as a consequence of the random dopant distribution. The conductance map Fig.~\ref{Fig2}a for sample A reveals more than 140 Coulomb diamonds measured at low gate bias above the mobility edge. The corresponding histogram of level spacings is shown Fig.~\ref{Fig2}b. Because the precision on the determination of the level spacing d$\delta \simeq 0.2$  meV is smaller than the width $\sim$~4.2 meV of the histogram, this broad histogram clearly indicates fluctuating level spacing in the nanowire. In this metallic regime, the level spacing distribution is described by random matrix theory\cite{Alhassid2000-cd,Beenakker1997-pg}, which predicts a bimodal Wigner distribution. This distribution is characterized by two maxima. One is at the Coulomb energy $E_{\rm C}$, the other at $E_{\rm C}+\langle\delta\rangle$, where $\langle\delta\rangle$= 1 meV is the mean level spacing calculated from the relation $\langle\delta\rangle = 1/\rho(E_F)V$ where $\rho(E_F)$ is the density of states, $E_F= 196$~meV is the Fermi energy corresponding to the estimated carrier density and $V$ is the volume of the QDot. The position of these two peaks are indicated on the histogram Fig.~2b.

In the random matrix theory, the random distribution of level spacings is accompanied by a random distribution of the amplitude of the wavefunctions at the boundaries of the QDot, which leads to a broad distribution of the conductance of the Coulomb peak maxima, as observed experimentally and shown in Fig.~\ref{Fig2}c. This last observation implies strong fluctuations in the coupling $\Gamma_{\rm S}$ of the levels to the superconducting electrodes which has important consequence on the formation of the superconducting SGRs. Finally, from the magnetic field dependence of the Zeeman splitting, Fig. S3, a factor $|$g$|\simeq 10$ is extracted, which is consistent with known values of the g-factor in InAs\cite{Csonka2008-jy}.

\begin{figure}[ht!]
	\begin{center}
		\includegraphics[width=7cm]{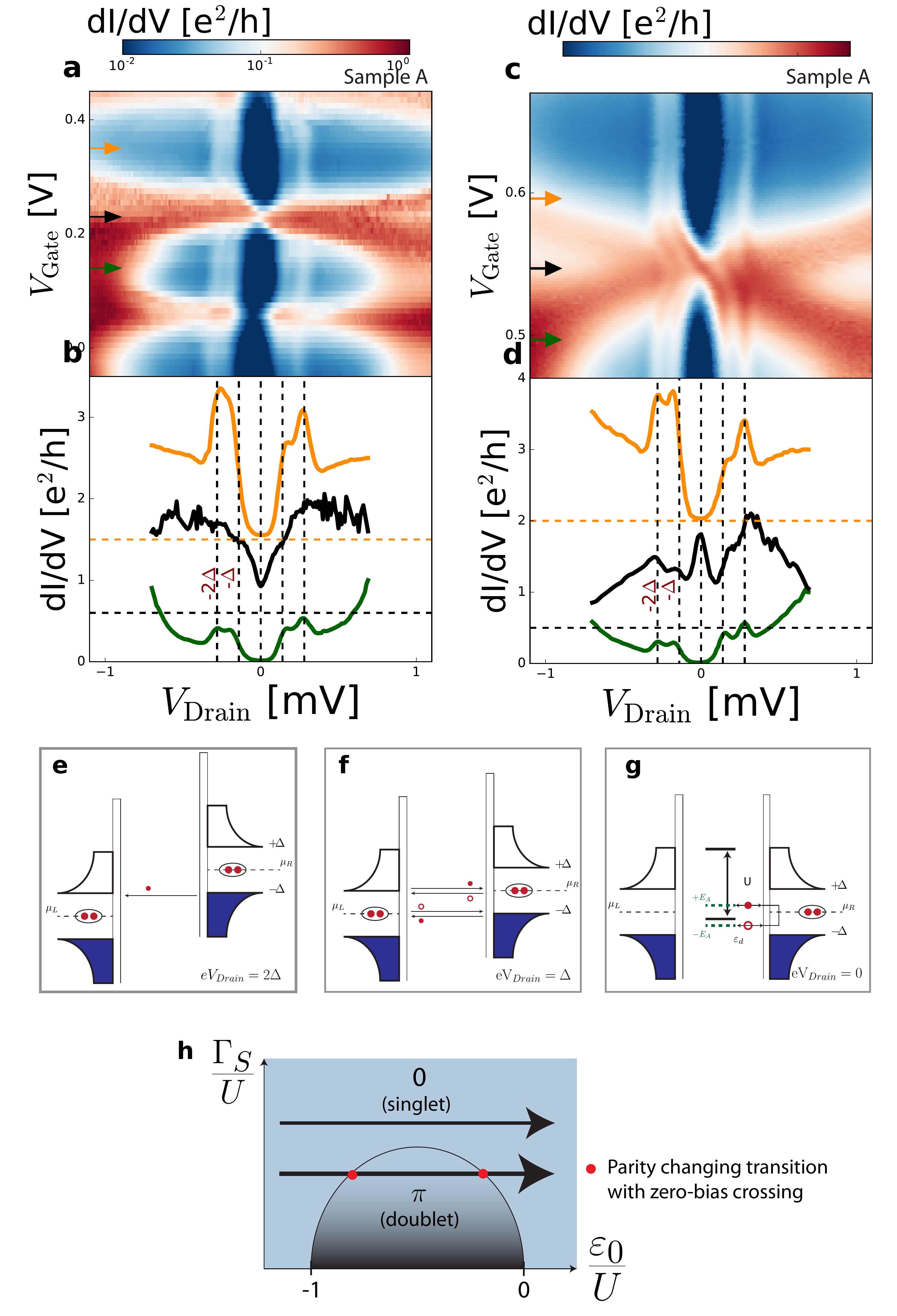}
		\caption{\label{Fig3} The conductance maps a) c) and the corresponding plot at selected gate voltages b) d) show the spectra of the Andreev QDot for two regimes of weak and strong coupling $\Gamma_{\rm S}$, respectively. In the strong coupling regime, only the conductance ridges at $2\Delta/e$ and $\Delta/e$ are seen. They are respectively due to the peak in the density of states at the gap edge, panel e), and to the first order multiple Andreev reflection (n=1), panel f). In the weak coupling regime, SGRs form at energies below the superconducting gap energy, panel g). The parity changing phase transition is identified by a zero bias crossing of the SGRs.
		h) Phase diagram  of the S-QD-S system as function of the charging energy $U$, the QD level $\varepsilon_0$ and the coupling to the lead $\Gamma_{\rm S}$. At weak coupling $\Gamma_{\rm S}$, only the singlet ground state can be observed upon changing the gate voltage. At strong coupling, a parity changing phase transition from the singlet phase to the doublet phase is identified as a zero-crossing of the SGRs.}
	\end{center}
\end{figure}

Figure~\ref{Fig3}a,b show conductance maps measured at low bias. It reveals the superconducting gap of Al with 4 lines symmetric about zero bias. Inside the Coulomb diamonds, the horizontal conductance ridges at eV$_{\rm Drain}$ = 2$\Delta$ correspond to the onset of elastic quasiparticle co-tunneling where the peaks in the density of states of the two superconducting electrodes are aligned, as sketched Fig.~\ref{Fig3}e. The superconducting gap value $\Delta\simeq 140 \mu$V is extracted, which is similar to values obtained in past works\cite{Buizert2007-bn}. This figure also shows conductance ridges at eV$_{\rm Drain} = \Delta$. They correspond to n=1 Andreev reflections as depicted Fig.~\ref{Fig3}h. They are generally expected in SNS structures\cite{Octavio1983-bj,Martin-Rodero2011-ji}, where the coupling to the electrodes is large, which corresponds to the singlet '0' regime in the phase diagram, Fig.\ref{Fig3}h. They have been observed previously in InAs nanowires\cite{Sand-Jespersen2007-zp} and carbon nanotubes\cite{Buitelaar2003-ae,Eichler2007-lq,Grove-Rasmussen2009-jb,Pillet2010-wt}. 
For certain gate voltages, the pair of Andreev conductance peaks crosses at zero-bias, as shown Fig.~\ref{Fig3}c and Fig.S2. This zero-bias crossing is the consequence of a quantum phase transition with parity change of the S-QDot-S ground state\cite{Martin-Rodero2011-ji} from the singlet '0' state to the doublet '$\pi$' state. This transition is expected at low coupling $\Gamma_{\rm S}$ as indicated on the phase diagram Fig.~\ref{Fig3}h. The origin of these occasional zero-bias crossing is the consequence of fluctuating coupling to the electrodes as discussed above. Only those levels which are weakly coupled will show a zero-bias crossing. All these superconducting-like features disappear by the application of a small magnetic field about 20 mT.




\begin{figure}[ht!]
	\begin{center}
		\includegraphics[width=7cm]{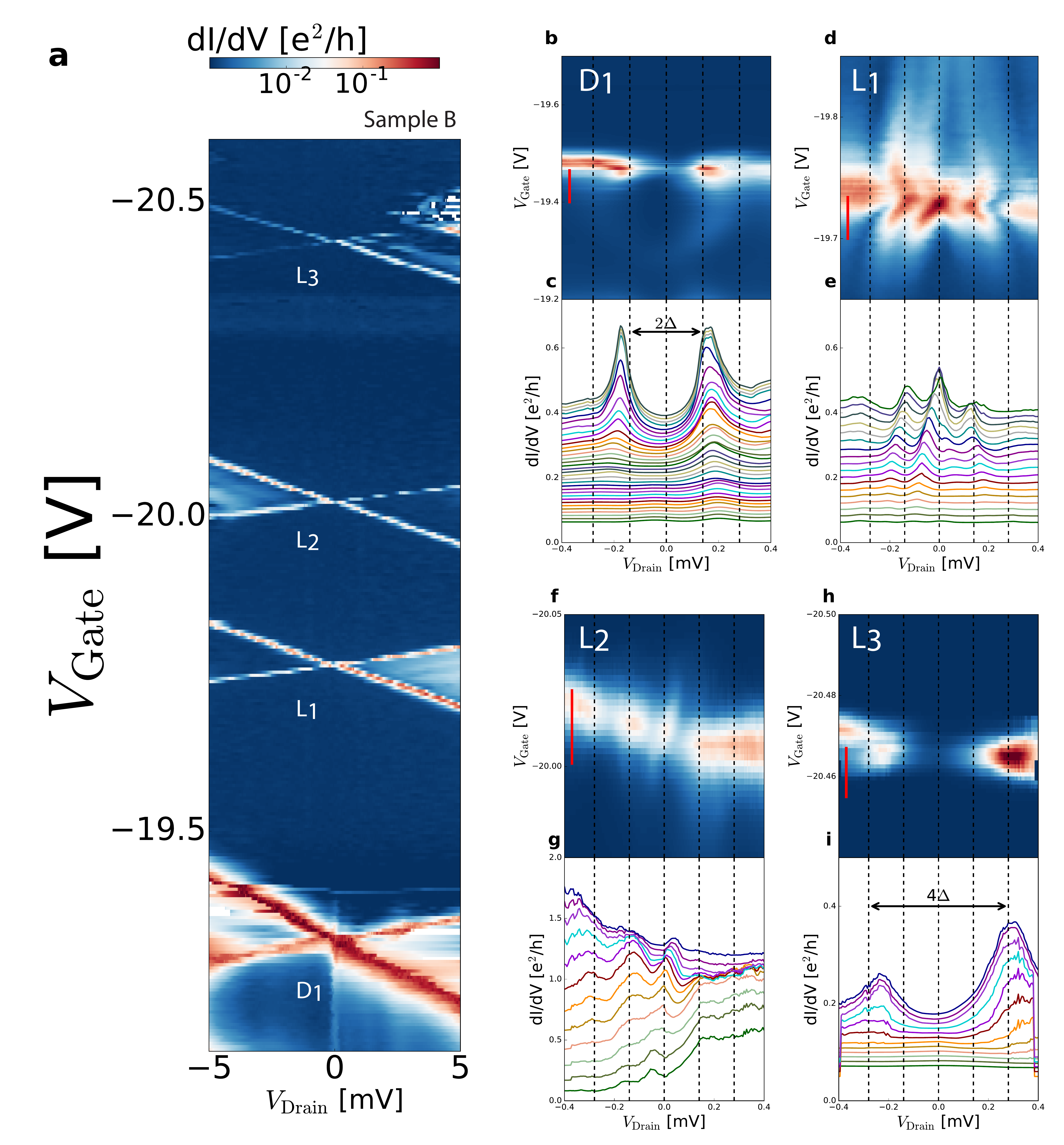}
		\caption{\label{Fig4} Conductance map a) and corresponding zooms b,d,f,h) showing the last delocalized level (D$_1$) and the localized impurity levels (L$_1$,L$_2$,L$_3$). The waterfall plots are shown for the range indicated by vertical red bars on the corresponding color maps. The shifts between curves are respectively of 0.01, 0.02, 0.1,  0.01 $e^2/h$.
		On the level D$_1$, b)c), only conductance peaks at $\Delta$ are observed, which are signature of Andreev reflections (n=1). On the first L$_1$, d) e), and second L$_2$, f) g), localized levels, a crossing of the conductance peak at zero bias is observed, which is the signature of the formation of the Shiba state. Finally, on the last localized level L$_3$, only the conductance peaks at 2$\Delta$ are observed, indicating the absence of Andreev reflections and only the presence of quasiparticle co-tunneling.}
	\end{center}
\end{figure}

We now turn to the evolution of the SGRs across the mobility edge. As a consequence of the decreasing coupling $\Gamma_{\rm S}$ of the levels with the superconducting electrodes, the high resolution zoom, Fig.~\ref{Fig4}, shows a remarkable evolution of the SGRs spectrum from the last delocalized level D$_1$ to the localized levels L$_1$ to L$_3$. The level D$_1$ shows mostly conductance ridges at $\Delta$. This is the singlet '0' regime dominated by the n=1 Andreev reflections that occur at strong coupling. The first and second localized levels, L$_1$ and L$_2$, show a crossing of the SGRs at zero-energy. This crossing indicates a parity change and the formation of a Shiba resonance at the Si impurity carrying a single electron spin; this is the doublet '$\pi$' regime that occurs at weak coupling in the phase diagram Fig.~\ref{Fig3}g. Finally, the last level L$_3$ shows only the superconducting gap features at 2$\Delta$, which are the signatures of quasiparticle co-tunneling. This last level is so localized and weakly coupled to the electrodes that no Andreev reflection occurs in the QDot.


To summarize, while past works on InAs based Andreev QDots employed large diameter nanowires ($>60$~nm), we found that Andreev QDots could be fabricated with smaller (30 nm) diameter nanowires provided that they are initially doped to high carrier concentration of $2\times10^{18}$ cm$^{-3}$. We found that these nanowires could be driven across an Anderson MIT upon applying a large negative back gate voltage and we observed a remarkable evolution of the superconducting SGRs across the mobility edge as consequence of the rapidly changing coupling of the levels with the superconducting electrodes. Deeply localized levels do not allow for the presence of Andreev reflections or the formation of Shiba states and only quasiparticles co-tunneling is observed. For localized levels near the mobility edge, Shiba bound states form and a parity changing quantum phase transition is identified by a crossing of the SGRs at zero energy. Finally, just above the mobility edge, simple n=1 Andreev resonances are observed. Further studies of SGRs in disordered QDots could provide insight into the Anderson MIT, which remains a subject of intense theoretical interest\cite{Belitz1994-vn,Evers2008-jz}.





\bibliography{Bibliography}

\begin{thebibliography}{59}%
\makeatletter
\providecommand \@ifxundefined [1]{%
 \@ifx{#1\undefined}
}%
\providecommand \@ifnum [1]{%
 \ifnum #1\expandafter \@firstoftwo
 \else \expandafter \@secondoftwo
 \fi
}%
\providecommand \@ifx [1]{%
 \ifx #1\expandafter \@firstoftwo
 \else \expandafter \@secondoftwo
 \fi
}%
\providecommand \natexlab [1]{#1}%
\providecommand \enquote  [1]{``#1''}%
\providecommand \bibnamefont  [1]{#1}%
\providecommand \bibfnamefont [1]{#1}%
\providecommand \citenamefont [1]{#1}%
\providecommand \href@noop [0]{\@secondoftwo}%
\providecommand \href [0]{\begingroup \@sanitize@url \@href}%
\providecommand \@href[1]{\@@startlink{#1}\@@href}%
\providecommand \@@href[1]{\endgroup#1\@@endlink}%
\providecommand \@sanitize@url [0]{\catcode `\\12\catcode `\$12\catcode
  `\&12\catcode `\#12\catcode `\^12\catcode `\_12\catcode `\%12\relax}%
\providecommand \@@startlink[1]{}%
\providecommand \@@endlink[0]{}%
\providecommand \url  [0]{\begingroup\@sanitize@url \@url }%
\providecommand \@url [1]{\endgroup\@href {#1}{\urlprefix }}%
\providecommand \urlprefix  [0]{URL }%
\providecommand \Eprint [0]{\href }%
\providecommand \doibase [0]{http://dx.doi.org/}%
\providecommand \selectlanguage [0]{\@gobble}%
\providecommand \bibinfo  [0]{\@secondoftwo}%
\providecommand \bibfield  [0]{\@secondoftwo}%
\providecommand \translation [1]{[#1]}%
\providecommand \BibitemOpen [0]{}%
\providecommand \bibitemStop [0]{}%
\providecommand \bibitemNoStop [0]{.\EOS\space}%
\providecommand \EOS [0]{\spacefactor3000\relax}%
\providecommand \BibitemShut  [1]{\csname bibitem#1\endcsname}%
\let\auto@bib@innerbib\@empty
\bibitem [{\citenamefont {Rozhkov}\ and\ \citenamefont
  {Arovas}(1999)}]{Rozhkov1999-hd}%
  \BibitemOpen
  \bibfield  {author} {\bibinfo {author} {\bibfnamefont {A.~V.}\ \bibnamefont
  {Rozhkov}}\ and\ \bibinfo {author} {\bibfnamefont {D.~P.}\ \bibnamefont
  {Arovas}},\ }\href@noop {} {\bibfield  {journal} {\bibinfo  {journal} {Phys.
  Rev. Lett.}\ }\textbf {\bibinfo {volume} {82}},\ \bibinfo {pages} {2788}
  (\bibinfo {year} {1999})}\BibitemShut {NoStop}%
\bibitem [{\citenamefont {Mart{\'\i}n-Rodero}\ and\ \citenamefont
  {Yeyati}(2011)}]{Martin-Rodero2011-ji}%
  \BibitemOpen
  \bibfield  {author} {\bibinfo {author} {\bibfnamefont {A.}~\bibnamefont
  {Mart{\'\i}n-Rodero}}\ and\ \bibinfo {author} {\bibfnamefont {A.~L.}\
  \bibnamefont {Yeyati}},\ }\href@noop {} {\bibfield  {journal} {\bibinfo
  {journal} {Adv. Phys.}\ }\textbf {\bibinfo {volume} {60}},\ \bibinfo {pages}
  {899} (\bibinfo {year} {2011})}\BibitemShut {NoStop}%
\bibitem [{\citenamefont {Kulik}(1969)}]{Kulik1969-ho}%
  \BibitemOpen
  \bibfield  {author} {\bibinfo {author} {\bibfnamefont {I.~O.}\ \bibnamefont
  {Kulik}},\ }\href@noop {} {\bibfield  {journal} {\bibinfo  {journal} {Soviet
  Journal of Experimental and Theoretical Physics}\ }\textbf {\bibinfo {volume}
  {30}},\ \bibinfo {pages} {944} (\bibinfo {year} {1969})}\BibitemShut
  {NoStop}%
\bibitem [{\citenamefont {Octavio}\ \emph {et~al.}(1983)\citenamefont
  {Octavio}, \citenamefont {Tinkham}, \citenamefont {Blonder},\ and\
  \citenamefont {Klapwijk}}]{Octavio1983-bj}%
  \BibitemOpen
  \bibfield  {author} {\bibinfo {author} {\bibfnamefont {M.}~\bibnamefont
  {Octavio}}, \bibinfo {author} {\bibfnamefont {M.}~\bibnamefont {Tinkham}},
  \bibinfo {author} {\bibfnamefont {G.~E.}\ \bibnamefont {Blonder}}, \ and\
  \bibinfo {author} {\bibfnamefont {T.~M.}\ \bibnamefont {Klapwijk}},\
  }\href@noop {} {\bibfield  {journal} {\bibinfo  {journal} {Phys. Rev. B:
  Condens. Matter Mater. Phys.}\ }\textbf {\bibinfo {volume} {27}},\ \bibinfo
  {pages} {6739} (\bibinfo {year} {1983})}\BibitemShut {NoStop}%
\bibitem [{\citenamefont {Scheer}\ \emph {et~al.}(1998)\citenamefont {Scheer},
  \citenamefont {Agra{\"\i}t}, \citenamefont {Cuevas}, \citenamefont {Yeyati},
  \citenamefont {Ludoph}, \citenamefont {Mart{\'\i}n-Rodero}, \citenamefont
  {Bollinger}, \citenamefont {van Ruitenbeek},\ and\ \citenamefont
  {Urbina}}]{Scheer1998-ga}%
  \BibitemOpen
  \bibfield  {author} {\bibinfo {author} {\bibfnamefont {E.}~\bibnamefont
  {Scheer}}, \bibinfo {author} {\bibfnamefont {N.}~\bibnamefont {Agra{\"\i}t}},
  \bibinfo {author} {\bibfnamefont {J.~C.}\ \bibnamefont {Cuevas}}, \bibinfo
  {author} {\bibfnamefont {A.~L.}\ \bibnamefont {Yeyati}}, \bibinfo {author}
  {\bibfnamefont {B.}~\bibnamefont {Ludoph}}, \bibinfo {author} {\bibfnamefont
  {A.}~\bibnamefont {Mart{\'\i}n-Rodero}}, \bibinfo {author} {\bibfnamefont
  {G.~R.}\ \bibnamefont {Bollinger}}, \bibinfo {author} {\bibfnamefont {J.~M.}\
  \bibnamefont {van Ruitenbeek}}, \ and\ \bibinfo {author} {\bibfnamefont
  {C.}~\bibnamefont {Urbina}},\ }\href@noop {} {\bibfield  {journal} {\bibinfo
  {journal} {Nature}\ }\textbf {\bibinfo {volume} {394}},\ \bibinfo {pages}
  {154} (\bibinfo {year} {1998})}\BibitemShut {NoStop}%
\bibitem [{\citenamefont {van Dam}\ \emph {et~al.}(2006)\citenamefont {van
  Dam}, \citenamefont {Nazarov}, \citenamefont {Bakkers}, \citenamefont
  {De~Franceschi},\ and\ \citenamefont {Kouwenhoven}}]{Van_Dam2006-vi}%
  \BibitemOpen
  \bibfield  {author} {\bibinfo {author} {\bibfnamefont {J.~a.}\ \bibnamefont
  {van Dam}}, \bibinfo {author} {\bibfnamefont {Y.~V.}\ \bibnamefont
  {Nazarov}}, \bibinfo {author} {\bibfnamefont {E.~P. a.~M.}\ \bibnamefont
  {Bakkers}}, \bibinfo {author} {\bibfnamefont {S.}~\bibnamefont
  {De~Franceschi}}, \ and\ \bibinfo {author} {\bibfnamefont {L.~P.}\
  \bibnamefont {Kouwenhoven}},\ }\href@noop {} {\bibfield  {journal} {\bibinfo
  {journal} {Nature}\ }\textbf {\bibinfo {volume} {442}},\ \bibinfo {pages}
  {667} (\bibinfo {year} {2006})}\BibitemShut {NoStop}%
\bibitem [{\citenamefont {Cleuziou}\ \emph {et~al.}(2006)\citenamefont
  {Cleuziou}, \citenamefont {Wernsdorfer}, \citenamefont {Bouchiat},
  \citenamefont {Ondar{\c c}uhu},\ and\ \citenamefont
  {Monthioux}}]{Cleuziou2006-he}%
  \BibitemOpen
  \bibfield  {author} {\bibinfo {author} {\bibfnamefont {J.-P.}\ \bibnamefont
  {Cleuziou}}, \bibinfo {author} {\bibfnamefont {W.}~\bibnamefont
  {Wernsdorfer}}, \bibinfo {author} {\bibfnamefont {V.}~\bibnamefont
  {Bouchiat}}, \bibinfo {author} {\bibfnamefont {T.}~\bibnamefont {Ondar{\c
  c}uhu}}, \ and\ \bibinfo {author} {\bibfnamefont {M.}~\bibnamefont
  {Monthioux}},\ }\href@noop {} {\bibfield  {journal} {\bibinfo  {journal}
  {Nat. Nanotechnol.}\ }\textbf {\bibinfo {volume} {1}},\ \bibinfo {pages} {53}
  (\bibinfo {year} {2006})}\BibitemShut {NoStop}%
\bibitem [{\citenamefont {J{\o}rgensen}\ \emph {et~al.}(2007)\citenamefont
  {J{\o}rgensen}, \citenamefont {Novotn{\'y}}, \citenamefont {Grove-Rasmussen},
  \citenamefont {Flensberg},\ and\ \citenamefont
  {Lindelof}}]{Jorgensen2007-xc}%
  \BibitemOpen
  \bibfield  {author} {\bibinfo {author} {\bibfnamefont {H.~I.}\ \bibnamefont
  {J{\o}rgensen}}, \bibinfo {author} {\bibfnamefont {T.}~\bibnamefont
  {Novotn{\'y}}}, \bibinfo {author} {\bibfnamefont {K.}~\bibnamefont
  {Grove-Rasmussen}}, \bibinfo {author} {\bibfnamefont {K.}~\bibnamefont
  {Flensberg}}, \ and\ \bibinfo {author} {\bibfnamefont {P.~E.}\ \bibnamefont
  {Lindelof}},\ }\href@noop {} {\bibfield  {journal} {\bibinfo  {journal} {Nano
  Lett.}\ }\textbf {\bibinfo {volume} {7}},\ \bibinfo {pages} {2441} (\bibinfo
  {year} {2007})}\BibitemShut {NoStop}%
\bibitem [{\citenamefont {Maurand}\ \emph {et~al.}(2012)\citenamefont
  {Maurand}, \citenamefont {Meng}, \citenamefont {Bonet}, \citenamefont
  {Florens}, \citenamefont {Marty},\ and\ \citenamefont
  {Wernsdorfer}}]{Maurand2012-tb}%
  \BibitemOpen
  \bibfield  {author} {\bibinfo {author} {\bibfnamefont {R.}~\bibnamefont
  {Maurand}}, \bibinfo {author} {\bibfnamefont {T.}~\bibnamefont {Meng}},
  \bibinfo {author} {\bibfnamefont {E.}~\bibnamefont {Bonet}}, \bibinfo
  {author} {\bibfnamefont {S.}~\bibnamefont {Florens}}, \bibinfo {author}
  {\bibfnamefont {L.}~\bibnamefont {Marty}}, \ and\ \bibinfo {author}
  {\bibfnamefont {W.}~\bibnamefont {Wernsdorfer}},\ }\href@noop {} {\bibfield
  {journal} {\bibinfo  {journal} {Physical Review X}\ }\textbf {\bibinfo
  {volume} {2}},\ \bibinfo {pages} {011009} (\bibinfo {year}
  {2012})}\BibitemShut {NoStop}%
\bibitem [{\citenamefont {Kanai}\ \emph {et~al.}(2010)\citenamefont {Kanai},
  \citenamefont {Deacon}, \citenamefont {Oiwa}, \citenamefont {Yoshida},
  \citenamefont {Shibata}, \citenamefont {Hirakawa},\ and\ \citenamefont
  {Tarucha}}]{Kanai2010-wc}%
  \BibitemOpen
  \bibfield  {author} {\bibinfo {author} {\bibfnamefont {Y.}~\bibnamefont
  {Kanai}}, \bibinfo {author} {\bibfnamefont {R.~S.}\ \bibnamefont {Deacon}},
  \bibinfo {author} {\bibfnamefont {A.}~\bibnamefont {Oiwa}}, \bibinfo {author}
  {\bibfnamefont {K.}~\bibnamefont {Yoshida}}, \bibinfo {author} {\bibfnamefont
  {K.}~\bibnamefont {Shibata}}, \bibinfo {author} {\bibfnamefont
  {K.}~\bibnamefont {Hirakawa}}, \ and\ \bibinfo {author} {\bibfnamefont
  {S.}~\bibnamefont {Tarucha}},\ }\href@noop {} {\bibfield  {journal} {\bibinfo
   {journal} {Phys. Rev. B: Condens. Matter Mater. Phys.}\ }\textbf {\bibinfo
  {volume} {82}},\ \bibinfo {pages} {054512} (\bibinfo {year}
  {2010})}\BibitemShut {NoStop}%
\bibitem [{\citenamefont {Eichler}\ \emph {et~al.}(2007)\citenamefont
  {Eichler}, \citenamefont {Weiss}, \citenamefont {Oberholzer}, \citenamefont
  {Sch{\"o}nenberger}, \citenamefont {Levy~Yeyati}, \citenamefont {Cuevas},\
  and\ \citenamefont {Mart{\'\i}n-Rodero}}]{Eichler2007-lq}%
  \BibitemOpen
  \bibfield  {author} {\bibinfo {author} {\bibfnamefont {A.}~\bibnamefont
  {Eichler}}, \bibinfo {author} {\bibfnamefont {M.}~\bibnamefont {Weiss}},
  \bibinfo {author} {\bibfnamefont {S.}~\bibnamefont {Oberholzer}}, \bibinfo
  {author} {\bibfnamefont {C.}~\bibnamefont {Sch{\"o}nenberger}}, \bibinfo
  {author} {\bibfnamefont {A.}~\bibnamefont {Levy~Yeyati}}, \bibinfo {author}
  {\bibfnamefont {J.}~\bibnamefont {Cuevas}}, \ and\ \bibinfo {author}
  {\bibfnamefont {A.}~\bibnamefont {Mart{\'\i}n-Rodero}},\ }\href@noop {}
  {\bibfield  {journal} {\bibinfo  {journal} {Phys. Rev. Lett.}\ }\textbf
  {\bibinfo {volume} {99}},\ \bibinfo {pages} {126602} (\bibinfo {year}
  {2007})}\BibitemShut {NoStop}%
\bibitem [{\citenamefont {Sand-Jespersen}\ \emph {et~al.}(2007)\citenamefont
  {Sand-Jespersen}, \citenamefont {Paaske}, \citenamefont {Andersen},
  \citenamefont {Grove-Rasmussen}, \citenamefont {J{\o}rgensen}, \citenamefont
  {Aagesen}, \citenamefont {S{\o}rensen}, \citenamefont {Lindelof},
  \citenamefont {Flensberg},\ and\ \citenamefont
  {Nyg{\aa}rd}}]{Sand-Jespersen2007-zp}%
  \BibitemOpen
  \bibfield  {author} {\bibinfo {author} {\bibfnamefont {T.}~\bibnamefont
  {Sand-Jespersen}}, \bibinfo {author} {\bibfnamefont {J.}~\bibnamefont
  {Paaske}}, \bibinfo {author} {\bibfnamefont {B.}~\bibnamefont {Andersen}},
  \bibinfo {author} {\bibfnamefont {K.}~\bibnamefont {Grove-Rasmussen}},
  \bibinfo {author} {\bibfnamefont {H.}~\bibnamefont {J{\o}rgensen}}, \bibinfo
  {author} {\bibfnamefont {M.}~\bibnamefont {Aagesen}}, \bibinfo {author}
  {\bibfnamefont {C.}~\bibnamefont {S{\o}rensen}}, \bibinfo {author}
  {\bibfnamefont {P.}~\bibnamefont {Lindelof}}, \bibinfo {author}
  {\bibfnamefont {K.}~\bibnamefont {Flensberg}}, \ and\ \bibinfo {author}
  {\bibfnamefont {J.}~\bibnamefont {Nyg{\aa}rd}},\ }\href@noop {} {\bibfield
  {journal} {\bibinfo  {journal} {Phys. Rev. Lett.}\ }\textbf {\bibinfo
  {volume} {99}},\ \bibinfo {pages} {126603} (\bibinfo {year}
  {2007})}\BibitemShut {NoStop}%
\bibitem [{\citenamefont {Pillet}\ \emph {et~al.}(2010)\citenamefont {Pillet},
  \citenamefont {Quay}, \citenamefont {Morfin}, \citenamefont {Bena},
  \citenamefont {Yeyati},\ and\ \citenamefont {Joyez}}]{Pillet2010-wt}%
  \BibitemOpen
  \bibfield  {author} {\bibinfo {author} {\bibfnamefont {J.-D.}\ \bibnamefont
  {Pillet}}, \bibinfo {author} {\bibfnamefont {C.}~\bibnamefont {Quay}},
  \bibinfo {author} {\bibfnamefont {P.}~\bibnamefont {Morfin}}, \bibinfo
  {author} {\bibfnamefont {C.}~\bibnamefont {Bena}}, \bibinfo {author}
  {\bibfnamefont {a.~L.}\ \bibnamefont {Yeyati}}, \ and\ \bibinfo {author}
  {\bibfnamefont {P.}~\bibnamefont {Joyez}},\ }\href@noop {} {\bibfield
  {journal} {\bibinfo  {journal} {Nat. Phys.}\ }\textbf {\bibinfo {volume}
  {6}},\ \bibinfo {pages} {965} (\bibinfo {year} {2010})}\BibitemShut {NoStop}%
\bibitem [{\citenamefont {Deacon}\ \emph
  {et~al.}(2010{\natexlab{a}})\citenamefont {Deacon}, \citenamefont {Tanaka},
  \citenamefont {Oiwa}, \citenamefont {Sakano}, \citenamefont {Yoshida},
  \citenamefont {Shibata}, \citenamefont {Hirakawa},\ and\ \citenamefont
  {Tarucha}}]{Deacon2010-ub}%
  \BibitemOpen
  \bibfield  {author} {\bibinfo {author} {\bibfnamefont {R.~S.}\ \bibnamefont
  {Deacon}}, \bibinfo {author} {\bibfnamefont {Y.}~\bibnamefont {Tanaka}},
  \bibinfo {author} {\bibfnamefont {A.}~\bibnamefont {Oiwa}}, \bibinfo {author}
  {\bibfnamefont {R.}~\bibnamefont {Sakano}}, \bibinfo {author} {\bibfnamefont
  {K.}~\bibnamefont {Yoshida}}, \bibinfo {author} {\bibfnamefont
  {K.}~\bibnamefont {Shibata}}, \bibinfo {author} {\bibfnamefont
  {K.}~\bibnamefont {Hirakawa}}, \ and\ \bibinfo {author} {\bibfnamefont
  {S.}~\bibnamefont {Tarucha}},\ }\href@noop {} {\bibfield  {journal} {\bibinfo
   {journal} {Phys. Rev. Lett.}\ }\textbf {\bibinfo {volume} {104}},\ \bibinfo
  {pages} {076805} (\bibinfo {year} {2010}{\natexlab{a}})}\BibitemShut
  {NoStop}%
\bibitem [{\citenamefont {Giazotto}\ \emph {et~al.}(2011)\citenamefont
  {Giazotto}, \citenamefont {Spathis}, \citenamefont {Roddaro}, \citenamefont
  {Biswas}, \citenamefont {Taddei}, \citenamefont {Governale},\ and\
  \citenamefont {Sorba}}]{Giazotto2011-ra}%
  \BibitemOpen
  \bibfield  {author} {\bibinfo {author} {\bibfnamefont {F.}~\bibnamefont
  {Giazotto}}, \bibinfo {author} {\bibfnamefont {P.}~\bibnamefont {Spathis}},
  \bibinfo {author} {\bibfnamefont {S.}~\bibnamefont {Roddaro}}, \bibinfo
  {author} {\bibfnamefont {S.}~\bibnamefont {Biswas}}, \bibinfo {author}
  {\bibfnamefont {F.}~\bibnamefont {Taddei}}, \bibinfo {author} {\bibfnamefont
  {M.}~\bibnamefont {Governale}}, \ and\ \bibinfo {author} {\bibfnamefont
  {L.}~\bibnamefont {Sorba}},\ }\href@noop {} {\bibfield  {journal} {\bibinfo
  {journal} {Nat. Phys.}\ }\textbf {\bibinfo {volume} {7}},\ \bibinfo {pages}
  {857} (\bibinfo {year} {2011})}\BibitemShut {NoStop}%
\bibitem [{\citenamefont {Pillet}\ \emph {et~al.}(2013)\citenamefont {Pillet},
  \citenamefont {Joyez}, \citenamefont {{\v Z}itko},\ and\ \citenamefont
  {Goffman}}]{Pillet2013-lr}%
  \BibitemOpen
  \bibfield  {author} {\bibinfo {author} {\bibfnamefont {J.-D.}\ \bibnamefont
  {Pillet}}, \bibinfo {author} {\bibfnamefont {P.}~\bibnamefont {Joyez}},
  \bibinfo {author} {\bibfnamefont {R.}~\bibnamefont {{\v Z}itko}}, \ and\
  \bibinfo {author} {\bibfnamefont {M.~F.}\ \bibnamefont {Goffman}},\
  }\href@noop {} {\bibfield  {journal} {\bibinfo  {journal} {Phys. Rev. B:
  Condens. Matter Mater. Phys.}\ }\textbf {\bibinfo {volume} {88}},\ \bibinfo
  {pages} {045101} (\bibinfo {year} {2013})}\BibitemShut {NoStop}%
\bibitem [{\citenamefont {Kumar}\ \emph {et~al.}(2014)\citenamefont {Kumar},
  \citenamefont {Gaim}, \citenamefont {Steininger}, \citenamefont {Yeyati},
  \citenamefont {Mart{\'\i}n-Rodero}, \citenamefont {H{\"u}ttel},\ and\
  \citenamefont {Strunk}}]{Kumar2014-mx}%
  \BibitemOpen
  \bibfield  {author} {\bibinfo {author} {\bibfnamefont {A.}~\bibnamefont
  {Kumar}}, \bibinfo {author} {\bibfnamefont {M.}~\bibnamefont {Gaim}},
  \bibinfo {author} {\bibfnamefont {D.}~\bibnamefont {Steininger}}, \bibinfo
  {author} {\bibfnamefont {A.~L.}\ \bibnamefont {Yeyati}}, \bibinfo {author}
  {\bibfnamefont {A.}~\bibnamefont {Mart{\'\i}n-Rodero}}, \bibinfo {author}
  {\bibfnamefont {A.~K.}\ \bibnamefont {H{\"u}ttel}}, \ and\ \bibinfo {author}
  {\bibfnamefont {C.}~\bibnamefont {Strunk}},\ }\href@noop {} {\bibfield
  {journal} {\bibinfo  {journal} {Phys. Rev. B Condens. Matter}\ }\textbf
  {\bibinfo {volume} {89}},\ \bibinfo {pages} {075428} (\bibinfo {year}
  {2014})}\BibitemShut {NoStop}%
\bibitem [{\citenamefont {Dirks}\ \emph {et~al.}(2010)\citenamefont {Dirks},
  \citenamefont {Hughes}, \citenamefont {Lal}, \citenamefont {Uchoa},
  \citenamefont {Chen}, \citenamefont {Chialvo}, \citenamefont {Goldbart},\
  and\ \citenamefont {Mason}}]{Dirks2010-ll}%
  \BibitemOpen
  \bibfield  {author} {\bibinfo {author} {\bibfnamefont {T.}~\bibnamefont
  {Dirks}}, \bibinfo {author} {\bibfnamefont {T.~L.}\ \bibnamefont {Hughes}},
  \bibinfo {author} {\bibfnamefont {S.}~\bibnamefont {Lal}}, \bibinfo {author}
  {\bibfnamefont {B.}~\bibnamefont {Uchoa}}, \bibinfo {author} {\bibfnamefont
  {Y.-F.}\ \bibnamefont {Chen}}, \bibinfo {author} {\bibfnamefont
  {C.}~\bibnamefont {Chialvo}}, \bibinfo {author} {\bibfnamefont {P.~M.}\
  \bibnamefont {Goldbart}}, \ and\ \bibinfo {author} {\bibfnamefont
  {N.}~\bibnamefont {Mason}},\ }\href@noop {} {\bibfield  {journal} {\bibinfo
  {journal} {Nat. Phys.}\ }\textbf {\bibinfo {volume} {7}},\ \bibinfo {pages}
  {25} (\bibinfo {year} {2010})}\BibitemShut {NoStop}%
\bibitem [{\citenamefont {Deacon}\ \emph
  {et~al.}(2010{\natexlab{b}})\citenamefont {Deacon}, \citenamefont {Tanaka},
  \citenamefont {Oiwa}, \citenamefont {Sakano}, \citenamefont {Yoshida},
  \citenamefont {Shibata}, \citenamefont {Hirakawa},\ and\ \citenamefont
  {Tarucha}}]{Deacon2010-pq}%
  \BibitemOpen
  \bibfield  {author} {\bibinfo {author} {\bibfnamefont {R.~S.}\ \bibnamefont
  {Deacon}}, \bibinfo {author} {\bibfnamefont {Y.}~\bibnamefont {Tanaka}},
  \bibinfo {author} {\bibfnamefont {A.}~\bibnamefont {Oiwa}}, \bibinfo {author}
  {\bibfnamefont {R.}~\bibnamefont {Sakano}}, \bibinfo {author} {\bibfnamefont
  {K.}~\bibnamefont {Yoshida}}, \bibinfo {author} {\bibfnamefont
  {K.}~\bibnamefont {Shibata}}, \bibinfo {author} {\bibfnamefont
  {K.}~\bibnamefont {Hirakawa}}, \ and\ \bibinfo {author} {\bibfnamefont
  {S.}~\bibnamefont {Tarucha}},\ }\href@noop {} {\bibfield  {journal} {\bibinfo
   {journal} {Phys. Rev. B Condens. Matter}\ }\textbf {\bibinfo {volume}
  {81}},\ \bibinfo {pages} {121308} (\bibinfo {year}
  {2010}{\natexlab{b}})}\BibitemShut {NoStop}%
\bibitem [{\citenamefont {Lee}\ \emph {et~al.}(2012)\citenamefont {Lee},
  \citenamefont {Jiang}, \citenamefont {Aguado}, \citenamefont {Katsaros},
  \citenamefont {Lieber},\ and\ \citenamefont {De~Franceschi}}]{Lee2012-ew}%
  \BibitemOpen
  \bibfield  {author} {\bibinfo {author} {\bibfnamefont {E.~J.~H.}\
  \bibnamefont {Lee}}, \bibinfo {author} {\bibfnamefont {X.}~\bibnamefont
  {Jiang}}, \bibinfo {author} {\bibfnamefont {R.}~\bibnamefont {Aguado}},
  \bibinfo {author} {\bibfnamefont {G.}~\bibnamefont {Katsaros}}, \bibinfo
  {author} {\bibfnamefont {C.~M.}\ \bibnamefont {Lieber}}, \ and\ \bibinfo
  {author} {\bibfnamefont {S.}~\bibnamefont {De~Franceschi}},\ }\href@noop {}
  {\bibfield  {journal} {\bibinfo  {journal} {Phys. Rev. Lett.}\ }\textbf
  {\bibinfo {volume} {109}},\ \bibinfo {pages} {186802} (\bibinfo {year}
  {2012})}\BibitemShut {NoStop}%
\bibitem [{\citenamefont {Chang}\ \emph {et~al.}(2013)\citenamefont {Chang},
  \citenamefont {Manucharyan}, \citenamefont {Jespersen}, \citenamefont
  {Nyg{\aa}rd},\ and\ \citenamefont {Marcus}}]{Chang2013-wq}%
  \BibitemOpen
  \bibfield  {author} {\bibinfo {author} {\bibfnamefont {W.}~\bibnamefont
  {Chang}}, \bibinfo {author} {\bibfnamefont {V.~E.}\ \bibnamefont
  {Manucharyan}}, \bibinfo {author} {\bibfnamefont {T.~S.}\ \bibnamefont
  {Jespersen}}, \bibinfo {author} {\bibfnamefont {J.}~\bibnamefont
  {Nyg{\aa}rd}}, \ and\ \bibinfo {author} {\bibfnamefont {C.~M.}\ \bibnamefont
  {Marcus}},\ }\href@noop {} {\bibfield  {journal} {\bibinfo  {journal} {Phys.
  Rev. Lett.}\ }\textbf {\bibinfo {volume} {110}},\ \bibinfo {pages} {217005}
  (\bibinfo {year} {2013})}\BibitemShut {NoStop}%
\bibitem [{\citenamefont {Mourik}\ \emph {et~al.}(2012)\citenamefont {Mourik},
  \citenamefont {Zuo}, \citenamefont {Frolov}, \citenamefont {Plissard},
  \citenamefont {Bakkers},\ and\ \citenamefont {Kouwenhoven}}]{Mourik2012-ij}%
  \BibitemOpen
  \bibfield  {author} {\bibinfo {author} {\bibfnamefont {V.}~\bibnamefont
  {Mourik}}, \bibinfo {author} {\bibfnamefont {K.}~\bibnamefont {Zuo}},
  \bibinfo {author} {\bibfnamefont {S.~M.}\ \bibnamefont {Frolov}}, \bibinfo
  {author} {\bibfnamefont {S.~R.}\ \bibnamefont {Plissard}}, \bibinfo {author}
  {\bibfnamefont {E.~P. a.~M.}\ \bibnamefont {Bakkers}}, \ and\ \bibinfo
  {author} {\bibfnamefont {L.~P.}\ \bibnamefont {Kouwenhoven}},\ }\href@noop {}
  {\bibfield  {journal} {\bibinfo  {journal} {Science}\ }\textbf {\bibinfo
  {volume} {336}},\ \bibinfo {pages} {1003} (\bibinfo {year}
  {2012})}\BibitemShut {NoStop}%
\bibitem [{\citenamefont {Das}\ \emph {et~al.}(2012)\citenamefont {Das},
  \citenamefont {Ronen}, \citenamefont {Most}, \citenamefont {Oreg},
  \citenamefont {Heiblum},\ and\ \citenamefont {Shtrikman}}]{Das2012-hm}%
  \BibitemOpen
  \bibfield  {author} {\bibinfo {author} {\bibfnamefont {A.}~\bibnamefont
  {Das}}, \bibinfo {author} {\bibfnamefont {Y.}~\bibnamefont {Ronen}}, \bibinfo
  {author} {\bibfnamefont {Y.}~\bibnamefont {Most}}, \bibinfo {author}
  {\bibfnamefont {Y.}~\bibnamefont {Oreg}}, \bibinfo {author} {\bibfnamefont
  {M.}~\bibnamefont {Heiblum}}, \ and\ \bibinfo {author} {\bibfnamefont
  {H.}~\bibnamefont {Shtrikman}},\ }\href@noop {} {\bibfield  {journal}
  {\bibinfo  {journal} {Nat. Phys.}\ }\textbf {\bibinfo {volume} {8}},\
  \bibinfo {pages} {887} (\bibinfo {year} {2012})}\BibitemShut {NoStop}%
\bibitem [{\citenamefont {Deng}\ \emph {et~al.}(2012)\citenamefont {Deng},
  \citenamefont {Yu}, \citenamefont {Huang}, \citenamefont {Larsson},
  \citenamefont {Caroff},\ and\ \citenamefont {Xu}}]{Deng2012-mb}%
  \BibitemOpen
  \bibfield  {author} {\bibinfo {author} {\bibfnamefont {M.~T.}\ \bibnamefont
  {Deng}}, \bibinfo {author} {\bibfnamefont {C.~L.}\ \bibnamefont {Yu}},
  \bibinfo {author} {\bibfnamefont {G.~Y.}\ \bibnamefont {Huang}}, \bibinfo
  {author} {\bibfnamefont {M.}~\bibnamefont {Larsson}}, \bibinfo {author}
  {\bibfnamefont {P.}~\bibnamefont {Caroff}}, \ and\ \bibinfo {author}
  {\bibfnamefont {H.~Q.}\ \bibnamefont {Xu}},\ }\href@noop {} {\bibfield
  {journal} {\bibinfo  {journal} {Nano Lett.}\ }\textbf {\bibinfo {volume}
  {12}},\ \bibinfo {pages} {6414} (\bibinfo {year} {2012})}\BibitemShut
  {NoStop}%
\bibitem [{\citenamefont {Yu}(1965)}]{Yu1965-bp}%
  \BibitemOpen
  \bibfield  {author} {\bibinfo {author} {\bibfnamefont {L.}~\bibnamefont
  {Yu}},\ }\href@noop {} {\bibfield  {journal} {\bibinfo  {journal} {Acta Phys.
  Chim. Sin.}\ } (\bibinfo {year} {1965})}\BibitemShut {NoStop}%
\bibitem [{\citenamefont {Shiba}(1968)}]{Shiba1968-xa}%
  \BibitemOpen
  \bibfield  {author} {\bibinfo {author} {\bibfnamefont {H.}~\bibnamefont
  {Shiba}},\ }\href@noop {} {\bibfield  {journal} {\bibinfo  {journal} {Progr.
  Theoret. Phys.}\ }\textbf {\bibinfo {volume} {40}},\ \bibinfo {pages} {435}
  (\bibinfo {year} {1968})}\BibitemShut {NoStop}%
\bibitem [{\citenamefont {Rusinov}(1969)}]{Rusinov1969-fa}%
  \BibitemOpen
  \bibfield  {author} {\bibinfo {author} {\bibfnamefont {A.~I.}\ \bibnamefont
  {Rusinov}},\ }\href@noop {} {\bibfield  {journal} {\bibinfo  {journal}
  {SOVIET PHYSICS JETP}\ }\textbf {\bibinfo {volume} {29}} (\bibinfo {year}
  {1969})}\BibitemShut {NoStop}%
\bibitem [{\citenamefont {Soda}\ \emph {et~al.}(1967)\citenamefont {Soda},
  \citenamefont {Matsuura},\ and\ \citenamefont {Nagaoka}}]{Soda1967-dt}%
  \BibitemOpen
  \bibfield  {author} {\bibinfo {author} {\bibfnamefont {T.}~\bibnamefont
  {Soda}}, \bibinfo {author} {\bibfnamefont {T.}~\bibnamefont {Matsuura}}, \
  and\ \bibinfo {author} {\bibfnamefont {Y.}~\bibnamefont {Nagaoka}},\
  }\href@noop {} {\bibfield  {journal} {\bibinfo  {journal} {Progr. Theoret.
  Phys.}\ } (\bibinfo {year} {1967})}\BibitemShut {NoStop}%
\bibitem [{\citenamefont {Shiba}\ and\ \citenamefont
  {Soda}(1969)}]{Shiba1969-zs}%
  \BibitemOpen
  \bibfield  {author} {\bibinfo {author} {\bibfnamefont {H.}~\bibnamefont
  {Shiba}}\ and\ \bibinfo {author} {\bibfnamefont {T.}~\bibnamefont {Soda}},\
  }\href@noop {} {\bibfield  {journal} {\bibinfo  {journal} {Progr. Theoret.
  Phys.}\ }\textbf {\bibinfo {volume} {41}},\ \bibinfo {pages} {25} (\bibinfo
  {year} {1969})}\BibitemShut {NoStop}%
\bibitem [{\citenamefont {Sakurai}(1970)}]{Sakurai1970-ye}%
  \BibitemOpen
  \bibfield  {author} {\bibinfo {author} {\bibfnamefont {A.}~\bibnamefont
  {Sakurai}},\ }\href@noop {} {\bibfield  {journal} {\bibinfo  {journal}
  {Progr. Theoret. Phys.}\ }\textbf {\bibinfo {volume} {44}},\ \bibinfo {pages}
  {1472} (\bibinfo {year} {1970})}\BibitemShut {NoStop}%
\bibitem [{\citenamefont {Satori}\ \emph {et~al.}(1992)\citenamefont {Satori},
  \citenamefont {Shiba}, \citenamefont {Sakai},\ and\ \citenamefont
  {Shimizu}}]{Satori1992-uc}%
  \BibitemOpen
  \bibfield  {author} {\bibinfo {author} {\bibfnamefont {K.}~\bibnamefont
  {Satori}}, \bibinfo {author} {\bibfnamefont {H.}~\bibnamefont {Shiba}},
  \bibinfo {author} {\bibfnamefont {O.}~\bibnamefont {Sakai}}, \ and\ \bibinfo
  {author} {\bibfnamefont {Y.}~\bibnamefont {Shimizu}},\ }\href@noop {}
  {\bibfield  {journal} {\bibinfo  {journal} {J. Phys. Soc. Jpn.}\ }\textbf
  {\bibinfo {volume} {61}},\ \bibinfo {pages} {3239} (\bibinfo {year}
  {1992})}\BibitemShut {NoStop}%
\bibitem [{\citenamefont {Simonin}\ and\ \citenamefont
  {Allub}(1995)}]{Simonin1995-rk}%
  \BibitemOpen
  \bibfield  {author} {\bibinfo {author} {\bibfnamefont {J.}~\bibnamefont
  {Simonin}}\ and\ \bibinfo {author} {\bibfnamefont {R.}~\bibnamefont
  {Allub}},\ }\href@noop {} {\bibfield  {journal} {\bibinfo  {journal} {Phys.
  Rev. Lett.}\ }\textbf {\bibinfo {volume} {74}},\ \bibinfo {pages} {466}
  (\bibinfo {year} {1995})}\BibitemShut {NoStop}%
\bibitem [{\citenamefont {Salkola}\ \emph {et~al.}(1997)\citenamefont
  {Salkola}, \citenamefont {Balatsky},\ and\ \citenamefont
  {Schrieffer}}]{Salkola1997-ey}%
  \BibitemOpen
  \bibfield  {author} {\bibinfo {author} {\bibfnamefont {M.~I.}\ \bibnamefont
  {Salkola}}, \bibinfo {author} {\bibfnamefont {a.~V.}\ \bibnamefont
  {Balatsky}}, \ and\ \bibinfo {author} {\bibfnamefont {J.~R.}\ \bibnamefont
  {Schrieffer}},\ }\href@noop {} {\bibfield  {journal} {\bibinfo  {journal}
  {Phys. Rev. B: Condens. Matter Mater. Phys.}\ }\textbf {\bibinfo {volume}
  {55}},\ \bibinfo {pages} {12648} (\bibinfo {year} {1997})}\BibitemShut
  {NoStop}%
\bibitem [{\citenamefont {Balatsky}\ \emph {et~al.}(2006)\citenamefont
  {Balatsky}, \citenamefont {Vekhter},\ and\ \citenamefont
  {Zhu}}]{Balatsky2006-mp}%
  \BibitemOpen
  \bibfield  {author} {\bibinfo {author} {\bibfnamefont {A.}~\bibnamefont
  {Balatsky}}, \bibinfo {author} {\bibfnamefont {I.}~\bibnamefont {Vekhter}}, \
  and\ \bibinfo {author} {\bibfnamefont {J.-X.}\ \bibnamefont {Zhu}},\
  }\href@noop {} {\bibfield  {journal} {\bibinfo  {journal} {Rev. Mod. Phys.}\
  }\textbf {\bibinfo {volume} {78}},\ \bibinfo {pages} {373} (\bibinfo {year}
  {2006})}\BibitemShut {NoStop}%
\bibitem [{\citenamefont {Bauer}\ \emph {et~al.}(2007)\citenamefont {Bauer},
  \citenamefont {Oguri},\ and\ \citenamefont {Hewson}}]{Bauer2007-yn}%
  \BibitemOpen
  \bibfield  {author} {\bibinfo {author} {\bibfnamefont {J.}~\bibnamefont
  {Bauer}}, \bibinfo {author} {\bibfnamefont {A.}~\bibnamefont {Oguri}}, \ and\
  \bibinfo {author} {\bibfnamefont {A.~C.}\ \bibnamefont {Hewson}},\
  }\href@noop {} {\bibfield  {journal} {\bibinfo  {journal} {J. Phys. Condens.
  Matter}\ }\textbf {\bibinfo {volume} {19}},\ \bibinfo {pages} {486211}
  (\bibinfo {year} {2007})}\BibitemShut {NoStop}%
\bibitem [{\citenamefont {Tsai}\ \emph {et~al.}(2009)\citenamefont {Tsai},
  \citenamefont {Zhang}, \citenamefont {Fang},\ and\ \citenamefont
  {Hu}}]{Tsai2009-mk}%
  \BibitemOpen
  \bibfield  {author} {\bibinfo {author} {\bibfnamefont {W.-F.}\ \bibnamefont
  {Tsai}}, \bibinfo {author} {\bibfnamefont {Y.-Y.}\ \bibnamefont {Zhang}},
  \bibinfo {author} {\bibfnamefont {C.}~\bibnamefont {Fang}}, \ and\ \bibinfo
  {author} {\bibfnamefont {J.}~\bibnamefont {Hu}},\ }\href@noop {} {\bibfield
  {journal} {\bibinfo  {journal} {Phys. Rev. B Condens. Matter}\ }\textbf
  {\bibinfo {volume} {80}},\ \bibinfo {pages} {064513} (\bibinfo {year}
  {2009})}\BibitemShut {NoStop}%
\bibitem [{\citenamefont {Franke}\ \emph {et~al.}(2011)\citenamefont {Franke},
  \citenamefont {Schulze},\ and\ \citenamefont {Pascual}}]{Franke2011-fw}%
  \BibitemOpen
  \bibfield  {author} {\bibinfo {author} {\bibfnamefont {K.~J.}\ \bibnamefont
  {Franke}}, \bibinfo {author} {\bibfnamefont {G.}~\bibnamefont {Schulze}}, \
  and\ \bibinfo {author} {\bibfnamefont {J.~I.}\ \bibnamefont {Pascual}},\
  }\href@noop {} {\bibfield  {journal} {\bibinfo  {journal} {Science}\ }\textbf
  {\bibinfo {volume} {332}},\ \bibinfo {pages} {940} (\bibinfo {year}
  {2011})}\BibitemShut {NoStop}%
\bibitem [{\citenamefont {Yazdani}\ \emph {et~al.}(1999)\citenamefont
  {Yazdani}, \citenamefont {Howald}, \citenamefont {Lutz}, \citenamefont
  {Kapitulnik},\ and\ \citenamefont {Eigler}}]{Yazdani1999-ww}%
  \BibitemOpen
  \bibfield  {author} {\bibinfo {author} {\bibfnamefont {A.}~\bibnamefont
  {Yazdani}}, \bibinfo {author} {\bibfnamefont {C.}~\bibnamefont {Howald}},
  \bibinfo {author} {\bibfnamefont {C.}~\bibnamefont {Lutz}}, \bibinfo {author}
  {\bibfnamefont {A.}~\bibnamefont {Kapitulnik}}, \ and\ \bibinfo {author}
  {\bibfnamefont {D.}~\bibnamefont {Eigler}},\ }\href@noop {} {\bibfield
  {journal} {\bibinfo  {journal} {Phys. Rev. Lett.}\ }\textbf {\bibinfo
  {volume} {83}},\ \bibinfo {pages} {176} (\bibinfo {year} {1999})}\BibitemShut
  {NoStop}%
\bibitem [{\citenamefont {M{\'e}nard}\ \emph {et~al.}(2015)\citenamefont
  {M{\'e}nard}, \citenamefont {Guissart}, \citenamefont {Brun}, \citenamefont
  {Pons}, \citenamefont {Stolyarov}, \citenamefont {Debontridder},
  \citenamefont {Leclerc}, \citenamefont {Janod}, \citenamefont {Cario},
  \citenamefont {Roditchev}, \citenamefont {Simon},\ and\ \citenamefont
  {Cren}}]{Menard2015-yl}%
  \BibitemOpen
  \bibfield  {author} {\bibinfo {author} {\bibfnamefont {G.~C.}\ \bibnamefont
  {M{\'e}nard}}, \bibinfo {author} {\bibfnamefont {S.}~\bibnamefont
  {Guissart}}, \bibinfo {author} {\bibfnamefont {C.}~\bibnamefont {Brun}},
  \bibinfo {author} {\bibfnamefont {S.}~\bibnamefont {Pons}}, \bibinfo {author}
  {\bibfnamefont {V.~S.}\ \bibnamefont {Stolyarov}}, \bibinfo {author}
  {\bibfnamefont {F.}~\bibnamefont {Debontridder}}, \bibinfo {author}
  {\bibfnamefont {M.~V.}\ \bibnamefont {Leclerc}}, \bibinfo {author}
  {\bibfnamefont {E.}~\bibnamefont {Janod}}, \bibinfo {author} {\bibfnamefont
  {L.}~\bibnamefont {Cario}}, \bibinfo {author} {\bibfnamefont
  {D.}~\bibnamefont {Roditchev}}, \bibinfo {author} {\bibfnamefont
  {P.}~\bibnamefont {Simon}}, \ and\ \bibinfo {author} {\bibfnamefont
  {T.}~\bibnamefont {Cren}},\ }\href@noop {} {\bibfield  {journal} {\bibinfo
  {journal} {Nat. Phys.}\ ,\ \bibinfo {pages} {1}} (\bibinfo {year}
  {2015})}\BibitemShut {NoStop}%
\bibitem [{\citenamefont {Calvet}\ \emph {et~al.}(2002)\citenamefont {Calvet},
  \citenamefont {Wheeler},\ and\ \citenamefont {Reed}}]{Calvet2002-lp}%
  \BibitemOpen
  \bibfield  {author} {\bibinfo {author} {\bibfnamefont {L.~E.}\ \bibnamefont
  {Calvet}}, \bibinfo {author} {\bibfnamefont {R.~G.}\ \bibnamefont {Wheeler}},
  \ and\ \bibinfo {author} {\bibfnamefont {M.~a.}\ \bibnamefont {Reed}},\
  }\href@noop {} {\bibfield  {journal} {\bibinfo  {journal} {Appl. Phys.
  Lett.}\ }\textbf {\bibinfo {volume} {80}},\ \bibinfo {pages} {1761} (\bibinfo
  {year} {2002})}\BibitemShut {NoStop}%
\bibitem [{\citenamefont {Sellier}\ \emph {et~al.}(2006)\citenamefont
  {Sellier}, \citenamefont {Lansbergen}, \citenamefont {Caro}, \citenamefont
  {Rogge}, \citenamefont {Collaert}, \citenamefont {Ferain}, \citenamefont
  {Jurczak},\ and\ \citenamefont {Biesemans}}]{Sellier2006-ru}%
  \BibitemOpen
  \bibfield  {author} {\bibinfo {author} {\bibfnamefont {H.}~\bibnamefont
  {Sellier}}, \bibinfo {author} {\bibfnamefont {G.~P.}\ \bibnamefont
  {Lansbergen}}, \bibinfo {author} {\bibfnamefont {J.}~\bibnamefont {Caro}},
  \bibinfo {author} {\bibfnamefont {S.}~\bibnamefont {Rogge}}, \bibinfo
  {author} {\bibfnamefont {N.}~\bibnamefont {Collaert}}, \bibinfo {author}
  {\bibfnamefont {I.}~\bibnamefont {Ferain}}, \bibinfo {author} {\bibfnamefont
  {M.}~\bibnamefont {Jurczak}}, \ and\ \bibinfo {author} {\bibfnamefont
  {S.}~\bibnamefont {Biesemans}},\ }\href@noop {} {\bibfield  {journal}
  {\bibinfo  {journal} {Phys. Rev. Lett.}\ }\textbf {\bibinfo {volume} {97}},\
  \bibinfo {pages} {206805} (\bibinfo {year} {2006})}\BibitemShut {NoStop}%
\bibitem [{\citenamefont {Calvet}\ \emph {et~al.}(2011)\citenamefont {Calvet},
  \citenamefont {Snyder},\ and\ \citenamefont {Wernsdorfer}}]{Calvet2011-ew}%
  \BibitemOpen
  \bibfield  {author} {\bibinfo {author} {\bibfnamefont {L.~E.}\ \bibnamefont
  {Calvet}}, \bibinfo {author} {\bibfnamefont {J.~P.}\ \bibnamefont {Snyder}},
  \ and\ \bibinfo {author} {\bibfnamefont {W.}~\bibnamefont {Wernsdorfer}},\
  }\href@noop {} {\bibfield  {journal} {\bibinfo  {journal} {Phys. Rev. B:
  Condens. Matter Mater. Phys.}\ }\textbf {\bibinfo {volume} {83}},\ \bibinfo
  {pages} {205415} (\bibinfo {year} {2011})}\BibitemShut {NoStop}%
\bibitem [{\citenamefont {Calvet}\ \emph
  {et~al.}(2007{\natexlab{a}})\citenamefont {Calvet}, \citenamefont {Wheeler},\
  and\ \citenamefont {Reed}}]{Calvet2007-fz}%
  \BibitemOpen
  \bibfield  {author} {\bibinfo {author} {\bibfnamefont {L.}~\bibnamefont
  {Calvet}}, \bibinfo {author} {\bibfnamefont {R.}~\bibnamefont {Wheeler}}, \
  and\ \bibinfo {author} {\bibfnamefont {M.}~\bibnamefont {Reed}},\ }\href@noop
  {} {\bibfield  {journal} {\bibinfo  {journal} {Phys. Rev. Lett.}\ }\textbf
  {\bibinfo {volume} {98}},\ \bibinfo {pages} {096805} (\bibinfo {year}
  {2007}{\natexlab{a}})}\BibitemShut {NoStop}%
\bibitem [{\citenamefont {Calvet}\ \emph
  {et~al.}(2007{\natexlab{b}})\citenamefont {Calvet}, \citenamefont {Wheeler},\
  and\ \citenamefont {Reed}}]{Calvet2007-jm}%
  \BibitemOpen
  \bibfield  {author} {\bibinfo {author} {\bibfnamefont {L.}~\bibnamefont
  {Calvet}}, \bibinfo {author} {\bibfnamefont {R.}~\bibnamefont {Wheeler}}, \
  and\ \bibinfo {author} {\bibfnamefont {M.}~\bibnamefont {Reed}},\ }\href@noop
  {} {\bibfield  {journal} {\bibinfo  {journal} {Phys. Rev. B: Condens. Matter
  Mater. Phys.}\ }\textbf {\bibinfo {volume} {76}},\ \bibinfo {pages} {035319}
  (\bibinfo {year} {2007}{\natexlab{b}})}\BibitemShut {NoStop}%
\bibitem [{\citenamefont {Mottaghizadeh}\ \emph {et~al.}(2014)\citenamefont
  {Mottaghizadeh}, \citenamefont {Yu}, \citenamefont {Lang}, \citenamefont
  {Zimmers},\ and\ \citenamefont {Aubin}}]{Mottaghizadeh2014-hn}%
  \BibitemOpen
  \bibfield  {author} {\bibinfo {author} {\bibfnamefont {A.}~\bibnamefont
  {Mottaghizadeh}}, \bibinfo {author} {\bibfnamefont {Q.}~\bibnamefont {Yu}},
  \bibinfo {author} {\bibfnamefont {P.~L.}\ \bibnamefont {Lang}}, \bibinfo
  {author} {\bibfnamefont {A.}~\bibnamefont {Zimmers}}, \ and\ \bibinfo
  {author} {\bibfnamefont {H.}~\bibnamefont {Aubin}},\ }\href@noop {}
  {\bibfield  {journal} {\bibinfo  {journal} {Phys. Rev. Lett.}\ }\textbf
  {\bibinfo {volume} {112}},\ \bibinfo {pages} {066803} (\bibinfo {year}
  {2014})}\BibitemShut {NoStop}%
\bibitem [{\citenamefont {Lee}(1985)}]{Lee1985-pe}%
  \BibitemOpen
  \bibfield  {author} {\bibinfo {author} {\bibfnamefont {P.~A.}\ \bibnamefont
  {Lee}},\ }\href@noop {} {\bibfield  {journal} {\bibinfo  {journal} {Rev. Mod.
  Phys.}\ }\textbf {\bibinfo {volume} {57}},\ \bibinfo {pages} {287} (\bibinfo
  {year} {1985})}\BibitemShut {NoStop}%
\bibitem [{\citenamefont {Glas}\ \emph {et~al.}(2007)\citenamefont {Glas},
  \citenamefont {Harmand},\ and\ \citenamefont {Patriarche}}]{Glas2007-ch}%
  \BibitemOpen
  \bibfield  {author} {\bibinfo {author} {\bibfnamefont {F.}~\bibnamefont
  {Glas}}, \bibinfo {author} {\bibfnamefont {J.-C.}\ \bibnamefont {Harmand}}, \
  and\ \bibinfo {author} {\bibfnamefont {G.}~\bibnamefont {Patriarche}},\
  }\href@noop {} {\bibfield  {journal} {\bibinfo  {journal} {Phys. Rev. Lett.}\
  }\textbf {\bibinfo {volume} {99}},\ \bibinfo {pages} {146101} (\bibinfo
  {year} {2007})}\BibitemShut {NoStop}%
\bibitem [{\citenamefont {Scheffler}\ \emph {et~al.}(2009)\citenamefont
  {Scheffler}, \citenamefont {Nadj-Perge}, \citenamefont {Kouwenhoven},
  \citenamefont {Borgstr{\"o}m},\ and\ \citenamefont
  {Bakkers}}]{Scheffler2009-nm}%
  \BibitemOpen
  \bibfield  {author} {\bibinfo {author} {\bibfnamefont {M.}~\bibnamefont
  {Scheffler}}, \bibinfo {author} {\bibfnamefont {S.}~\bibnamefont
  {Nadj-Perge}}, \bibinfo {author} {\bibfnamefont {L.~P.}\ \bibnamefont
  {Kouwenhoven}}, \bibinfo {author} {\bibfnamefont {M.~T.}\ \bibnamefont
  {Borgstr{\"o}m}}, \ and\ \bibinfo {author} {\bibfnamefont {E.~P. a.~M.}\
  \bibnamefont {Bakkers}},\ }\href@noop {} {\bibfield  {journal} {\bibinfo
  {journal} {J. Appl. Phys.}\ }\textbf {\bibinfo {volume} {106}},\ \bibinfo
  {pages} {124303} (\bibinfo {year} {2009})}\BibitemShut {NoStop}%
\bibitem [{\citenamefont {Ihn}(2004)}]{Ihn2004-gn}%
  \BibitemOpen
  \bibfield  {author} {\bibinfo {author} {\bibfnamefont {T.}~\bibnamefont
  {Ihn}},\ }\href@noop {} {\emph {\bibinfo {title} {Electronic Quantum
  Transport in Mesoscopic Semiconductor Structures}}},\ Springer Tracts in
  Modern Physics\ (\bibinfo  {publisher} {Springer New York},\ \bibinfo {year}
  {2004})\BibitemShut {NoStop}%
\bibitem [{\citenamefont {Mott}(1990)}]{Mott1990-ga}%
  \BibitemOpen
  \bibfield  {author} {\bibinfo {author} {\bibfnamefont {N.}~\bibnamefont
  {Mott}},\ }\href@noop {} {\emph {\bibinfo {title} {{Metal-Insulator}
  Transitions}}}\ (\bibinfo  {publisher} {Taylor \& Francis},\ \bibinfo {year}
  {1990})\BibitemShut {NoStop}%
\bibitem [{\citenamefont {Schubert}(2015)}]{Schubert2015-ne}%
  \BibitemOpen
  \bibfield  {author} {\bibinfo {author} {\bibfnamefont {E.~F.}\ \bibnamefont
  {Schubert}},\ }\href@noop {} {\emph {\bibinfo {title} {Doping in {III-V}
  Semiconductors:}}}\ (\bibinfo  {publisher} {E. Fred Schubert},\ \bibinfo
  {year} {2015})\BibitemShut {NoStop}%
\bibitem [{\citenamefont {Alhassid}(2000)}]{Alhassid2000-cd}%
  \BibitemOpen
  \bibfield  {author} {\bibinfo {author} {\bibfnamefont {Y.}~\bibnamefont
  {Alhassid}},\ }\href@noop {} {\bibfield  {journal} {\bibinfo  {journal} {Rev.
  Mod. Phys.}\ }\textbf {\bibinfo {volume} {72}},\ \bibinfo {pages} {895}
  (\bibinfo {year} {2000})}\BibitemShut {NoStop}%
\bibitem [{\citenamefont {Beenakker}(1997)}]{Beenakker1997-pg}%
  \BibitemOpen
  \bibfield  {author} {\bibinfo {author} {\bibfnamefont {C.~W.~J.}\
  \bibnamefont {Beenakker}},\ }\href@noop {} {\bibfield  {journal} {\bibinfo
  {journal} {Rev. Mod. Phys.}\ }\textbf {\bibinfo {volume} {69}},\ \bibinfo
  {pages} {731} (\bibinfo {year} {1997})}\BibitemShut {NoStop}%
\bibitem [{\citenamefont {Csonka}\ \emph {et~al.}(2008)\citenamefont {Csonka},
  \citenamefont {Hofstetter}, \citenamefont {Freitag}, \citenamefont
  {Oberholzer}, \citenamefont {Sch{\"o}nenberger}, \citenamefont {Jespersen},
  \citenamefont {Aagesen},\ and\ \citenamefont {Nyg{\aa}rd}}]{Csonka2008-jy}%
  \BibitemOpen
  \bibfield  {author} {\bibinfo {author} {\bibfnamefont {S.}~\bibnamefont
  {Csonka}}, \bibinfo {author} {\bibfnamefont {L.}~\bibnamefont {Hofstetter}},
  \bibinfo {author} {\bibfnamefont {F.}~\bibnamefont {Freitag}}, \bibinfo
  {author} {\bibfnamefont {S.}~\bibnamefont {Oberholzer}}, \bibinfo {author}
  {\bibfnamefont {C.}~\bibnamefont {Sch{\"o}nenberger}}, \bibinfo {author}
  {\bibfnamefont {T.~S.}\ \bibnamefont {Jespersen}}, \bibinfo {author}
  {\bibfnamefont {M.}~\bibnamefont {Aagesen}}, \ and\ \bibinfo {author}
  {\bibfnamefont {J.}~\bibnamefont {Nyg{\aa}rd}},\ }\href@noop {} {\bibfield
  {journal} {\bibinfo  {journal} {Nano Lett.}\ }\textbf {\bibinfo {volume}
  {8}},\ \bibinfo {pages} {3932} (\bibinfo {year} {2008})}\BibitemShut
  {NoStop}%
\bibitem [{\citenamefont {Buizert}\ \emph {et~al.}(2007)\citenamefont
  {Buizert}, \citenamefont {Oiwa}, \citenamefont {Shibata}, \citenamefont
  {Hirakawa},\ and\ \citenamefont {Tarucha}}]{Buizert2007-bn}%
  \BibitemOpen
  \bibfield  {author} {\bibinfo {author} {\bibfnamefont {C.}~\bibnamefont
  {Buizert}}, \bibinfo {author} {\bibfnamefont {A.}~\bibnamefont {Oiwa}},
  \bibinfo {author} {\bibfnamefont {K.}~\bibnamefont {Shibata}}, \bibinfo
  {author} {\bibfnamefont {K.}~\bibnamefont {Hirakawa}}, \ and\ \bibinfo
  {author} {\bibfnamefont {S.}~\bibnamefont {Tarucha}},\ }\href@noop {}
  {\bibfield  {journal} {\bibinfo  {journal} {Phys. Rev. Lett.}\ }\textbf
  {\bibinfo {volume} {99}},\ \bibinfo {pages} {136806} (\bibinfo {year}
  {2007})}\BibitemShut {NoStop}%
\bibitem [{\citenamefont {Buitelaar}\ \emph {et~al.}(2003)\citenamefont
  {Buitelaar}, \citenamefont {Belzig}, \citenamefont {Nussbaumer},
  \citenamefont {Babi{\'c}}, \citenamefont {Bruder},\ and\ \citenamefont
  {Sch{\"o}nenberger}}]{Buitelaar2003-ae}%
  \BibitemOpen
  \bibfield  {author} {\bibinfo {author} {\bibfnamefont {M.}~\bibnamefont
  {Buitelaar}}, \bibinfo {author} {\bibfnamefont {W.}~\bibnamefont {Belzig}},
  \bibinfo {author} {\bibfnamefont {T.}~\bibnamefont {Nussbaumer}}, \bibinfo
  {author} {\bibfnamefont {B.}~\bibnamefont {Babi{\'c}}}, \bibinfo {author}
  {\bibfnamefont {C.}~\bibnamefont {Bruder}}, \ and\ \bibinfo {author}
  {\bibfnamefont {C.}~\bibnamefont {Sch{\"o}nenberger}},\ }\href@noop {}
  {\bibfield  {journal} {\bibinfo  {journal} {Phys. Rev. Lett.}\ }\textbf
  {\bibinfo {volume} {91}},\ \bibinfo {pages} {057005} (\bibinfo {year}
  {2003})}\BibitemShut {NoStop}%
\bibitem [{\citenamefont {Grove-Rasmussen}\ \emph {et~al.}(2009)\citenamefont
  {Grove-Rasmussen}, \citenamefont {J{\o}rgensen}, \citenamefont {Andersen},
  \citenamefont {Paaske}, \citenamefont {Jespersen}, \citenamefont
  {Nyg{\aa}rd}, \citenamefont {Flensberg},\ and\ \citenamefont
  {Lindelof}}]{Grove-Rasmussen2009-jb}%
  \BibitemOpen
  \bibfield  {author} {\bibinfo {author} {\bibfnamefont {K.}~\bibnamefont
  {Grove-Rasmussen}}, \bibinfo {author} {\bibfnamefont {H.~I.}\ \bibnamefont
  {J{\o}rgensen}}, \bibinfo {author} {\bibfnamefont {B.~M.}\ \bibnamefont
  {Andersen}}, \bibinfo {author} {\bibfnamefont {J.}~\bibnamefont {Paaske}},
  \bibinfo {author} {\bibfnamefont {T.~S.}\ \bibnamefont {Jespersen}}, \bibinfo
  {author} {\bibfnamefont {J.}~\bibnamefont {Nyg{\aa}rd}}, \bibinfo {author}
  {\bibfnamefont {K.}~\bibnamefont {Flensberg}}, \ and\ \bibinfo {author}
  {\bibfnamefont {P.~E.}\ \bibnamefont {Lindelof}},\ }\href@noop {} {\bibfield
  {journal} {\bibinfo  {journal} {Phys. Rev. B}\ }\textbf {\bibinfo {volume}
  {79}},\ \bibinfo {pages} {2} (\bibinfo {year} {2009})}\BibitemShut {NoStop}%
\bibitem [{\citenamefont {Belitz}\ and\ \citenamefont
  {Kirkpatrick}(1994)}]{Belitz1994-vn}%
  \BibitemOpen
  \bibfield  {author} {\bibinfo {author} {\bibfnamefont {D.}~\bibnamefont
  {Belitz}}\ and\ \bibinfo {author} {\bibfnamefont {T.}~\bibnamefont
  {Kirkpatrick}},\ }\href@noop {} {\bibfield  {journal} {\bibinfo  {journal}
  {Rev. Mod. Phys.}\ }\textbf {\bibinfo {volume} {66}},\ \bibinfo {pages} {261}
  (\bibinfo {year} {1994})}\BibitemShut {NoStop}%
\bibitem [{\citenamefont {Evers}\ and\ \citenamefont
  {Mirlin}(2008)}]{Evers2008-jz}%
  \BibitemOpen
  \bibfield  {author} {\bibinfo {author} {\bibfnamefont {F.}~\bibnamefont
  {Evers}}\ and\ \bibinfo {author} {\bibfnamefont {A.}~\bibnamefont {Mirlin}},\
  }\href@noop {} {\bibfield  {journal} {\bibinfo  {journal} {Rev. Mod. Phys.}\
  }\textbf {\bibinfo {volume} {80}},\ \bibinfo {pages} {1355} (\bibinfo {year}
  {2008})}\BibitemShut {NoStop}%
\end{thebibliography}%


\begin{thebibliography}{0}%
\makeatletter
\providecommand \@ifxundefined [1]{%
 \@ifx{#1\undefined}
}%
\providecommand \@ifnum [1]{%
 \ifnum #1\expandafter \@firstoftwo
 \else \expandafter \@secondoftwo
 \fi
}%
\providecommand \@ifx [1]{%
 \ifx #1\expandafter \@firstoftwo
 \else \expandafter \@secondoftwo
 \fi
}%
\providecommand \natexlab [1]{#1}%
\providecommand \enquote  [1]{``#1''}%
\providecommand \bibnamefont  [1]{#1}%
\providecommand \bibfnamefont [1]{#1}%
\providecommand \citenamefont [1]{#1}%
\providecommand \href@noop [0]{\@secondoftwo}%
\providecommand \href [0]{\begingroup \@sanitize@url \@href}%
\providecommand \@href[1]{\@@startlink{#1}\@@href}%
\providecommand \@@href[1]{\endgroup#1\@@endlink}%
\providecommand \@sanitize@url [0]{\catcode `\\12\catcode `\$12\catcode
  `\&12\catcode `\#12\catcode `\^12\catcode `\_12\catcode `\%12\relax}%
\providecommand \@@startlink[1]{}%
\providecommand \@@endlink[0]{}%
\providecommand \url  [0]{\begingroup\@sanitize@url \@url }%
\providecommand \@url [1]{\endgroup\@href {#1}{\urlprefix }}%
\providecommand \urlprefix  [0]{URL }%
\providecommand \Eprint [0]{\href }%
\providecommand \doibase [0]{http://dx.doi.org/}%
\providecommand \selectlanguage [0]{\@gobble}%
\providecommand \bibinfo  [0]{\@secondoftwo}%
\providecommand \bibfield  [0]{\@secondoftwo}%
\providecommand \translation [1]{[#1]}%
\providecommand \BibitemOpen [0]{}%
\providecommand \bibitemStop [0]{}%
\providecommand \bibitemNoStop [0]{.\EOS\space}%
\providecommand \EOS [0]{\spacefactor3000\relax}%
\providecommand \BibitemShut  [1]{\csname bibitem#1\endcsname}%
\let\auto@bib@innerbib\@empty
\end{thebibliography}%

\end{document}


\preprint{APS/123-QED}

\title{Shiba Bound States across the mobility edge in doped InAs nanowires}

\author{Alexandre Assouline}
\affiliation{LPEM, ESPCI Paris, PSL Research University; CNRS; Sorbonne Universit\'es, UPMC University of Paris 6\\
10 rue Vauquelin, F-75005 Paris, France}

\author{Cheryl Feuillet-Palma}
\affiliation{LPEM, ESPCI Paris, PSL Research University; CNRS; Sorbonne Universit\'es, UPMC University of Paris 6\\
10 rue Vauquelin, F-75005 Paris, France}

\author{Alexandre Zimmers}
\affiliation{LPEM, ESPCI Paris, PSL Research University; CNRS; Sorbonne Universit\'es, UPMC University of Paris 6\\
10 rue Vauquelin, F-75005 Paris, France}

\author{Marco Aprili}
\affiliation{Laboratoire de Physique des Solides, CNRS, Univ. Paris-Sud, University Paris-Saclay, 91405 Orsay Cedex, France}

\author{Jean-Christophe Harmand}
\affiliation{Centre de Nanosciences et de Nanotechnologies, CNRS, Univ. Paris-Sud, Universit\'es Paris-Saclay, C2N – Marcoussis, 91460 Marcoussis, France}

\author{Herv\'e Aubin}
\email{Herve.Aubin@espci.fr} 
\affiliation{LPEM, ESPCI Paris, PSL Research University; CNRS; Sorbonne Universit\'es, UPMC University of Paris 6\\
10 rue Vauquelin, F-75005 Paris, France}

\date{\today}

\pacs{Valid PACS appear here}
\maketitle

Additional figures.


\begin{figure*}[ht!]
	\begin{center}
		\includegraphics[width=16cm]{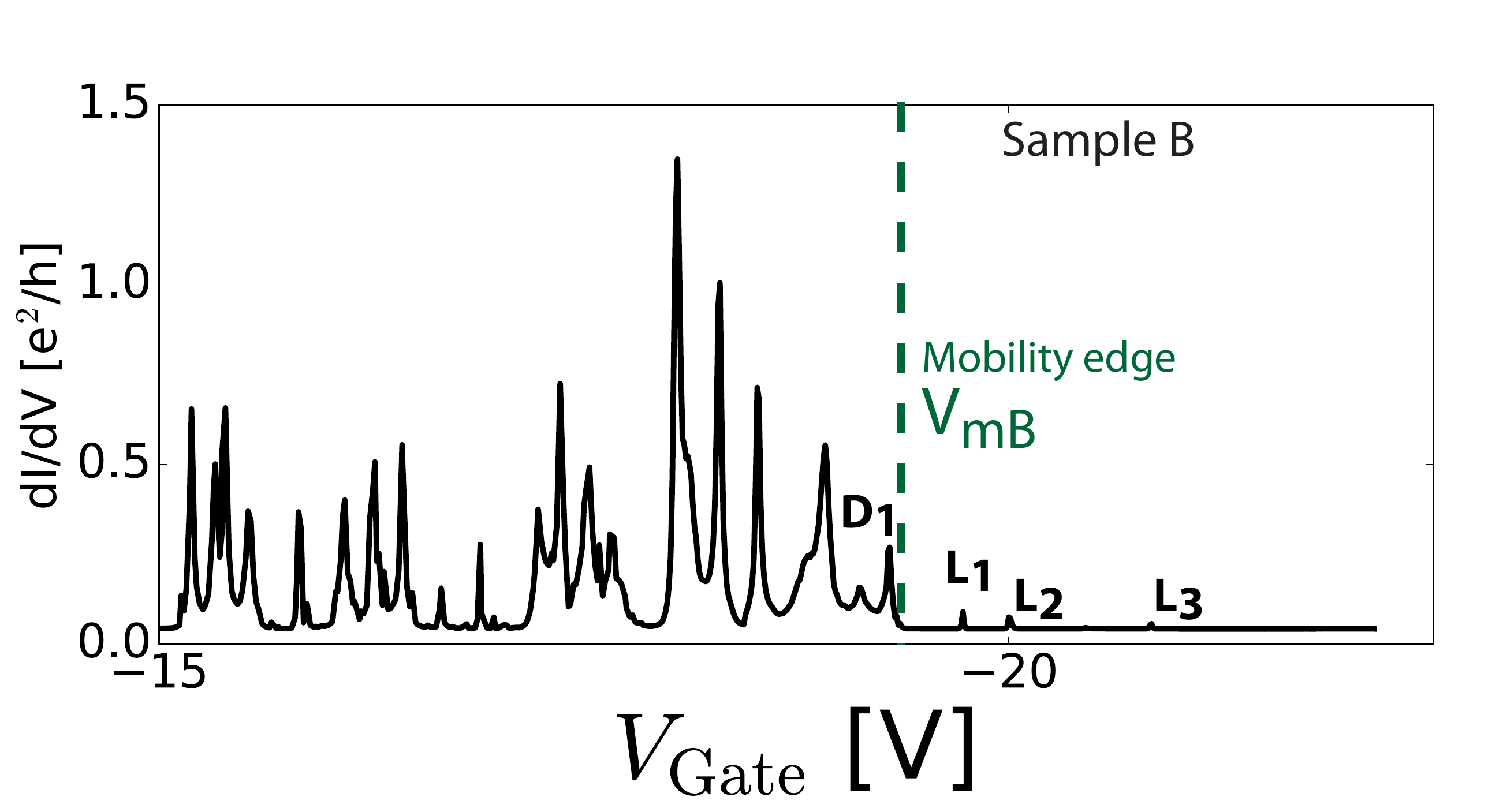}
		\caption{\label{FigS1} Zero-bias conductance as a function of back gate voltage for sample B on a large range than Fig.~1, showing that no other excitation levels are observed beyond L3, i.e. the nanowire is fully depleted.}
	\end{center}
\end{figure*}

\begin{figure*}[ht!]
	\begin{center}
		\includegraphics[width=16cm]{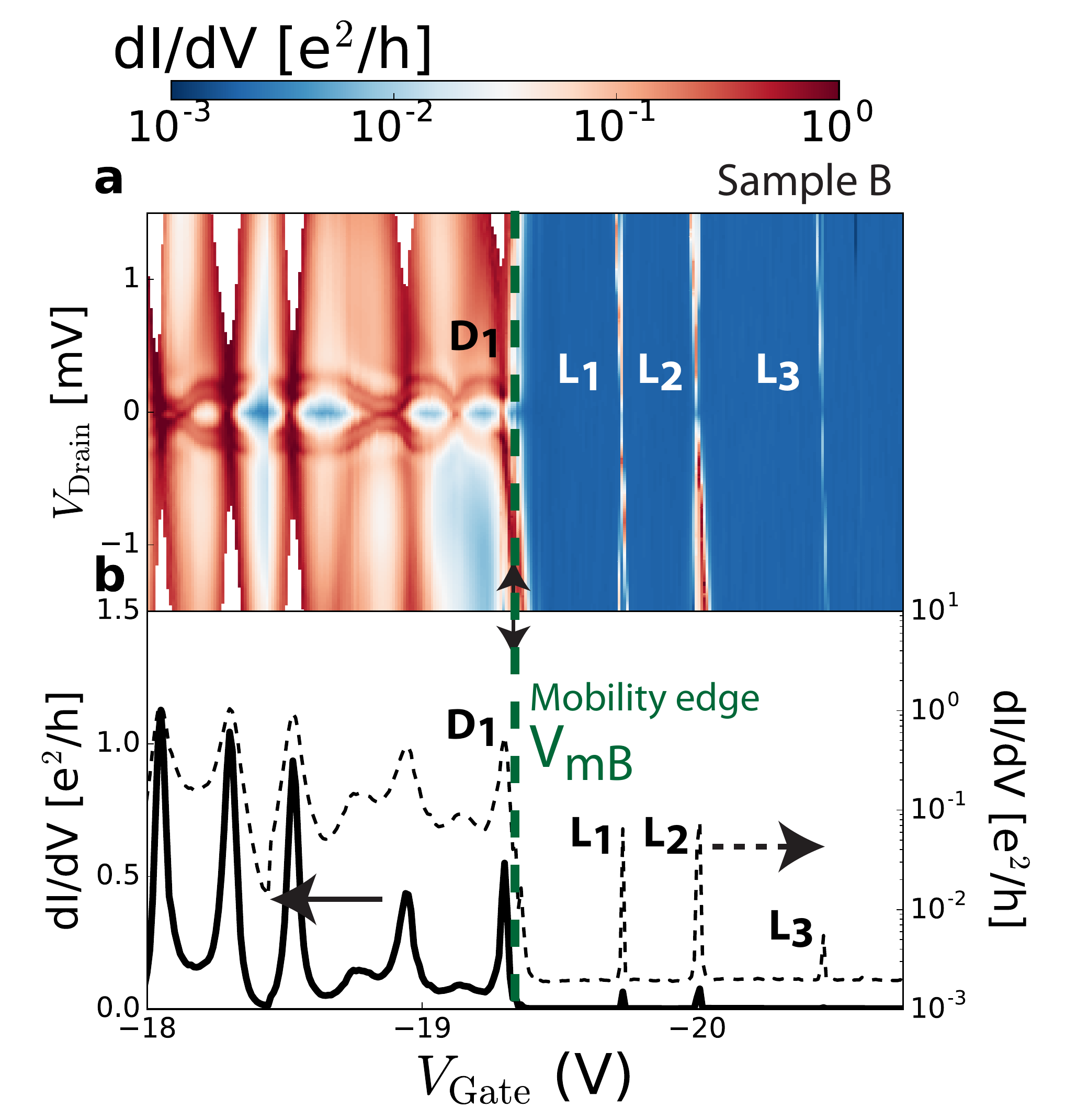}
		\caption{\label{FigS2} a) Conductance dI/dV maps for nanowire B as function of drain bias $V_{\rm Drain}$ and back-gate voltage $V_{\rm Gate}$ taken with a higher resolution on a smaller drain voltage  range than the main figure 1f. The map show the crossing of the mobility edge with the back-gate voltage below which only a few excited levels remain visible, labelled L$_1$ to L$_3$. The last delocalized level above the mobility edge is labelled D$_1$. Above this mobility edge, SGRs crossing zero-bias can also be observed. b) Zero-bias conductance as a function of back gate voltage plotted on linear scale (continuous line) and log-scale (dash line).}
	\end{center}
\end{figure*}

\begin{figure*}[ht!]
	\begin{center}
		\includegraphics[width=16cm]{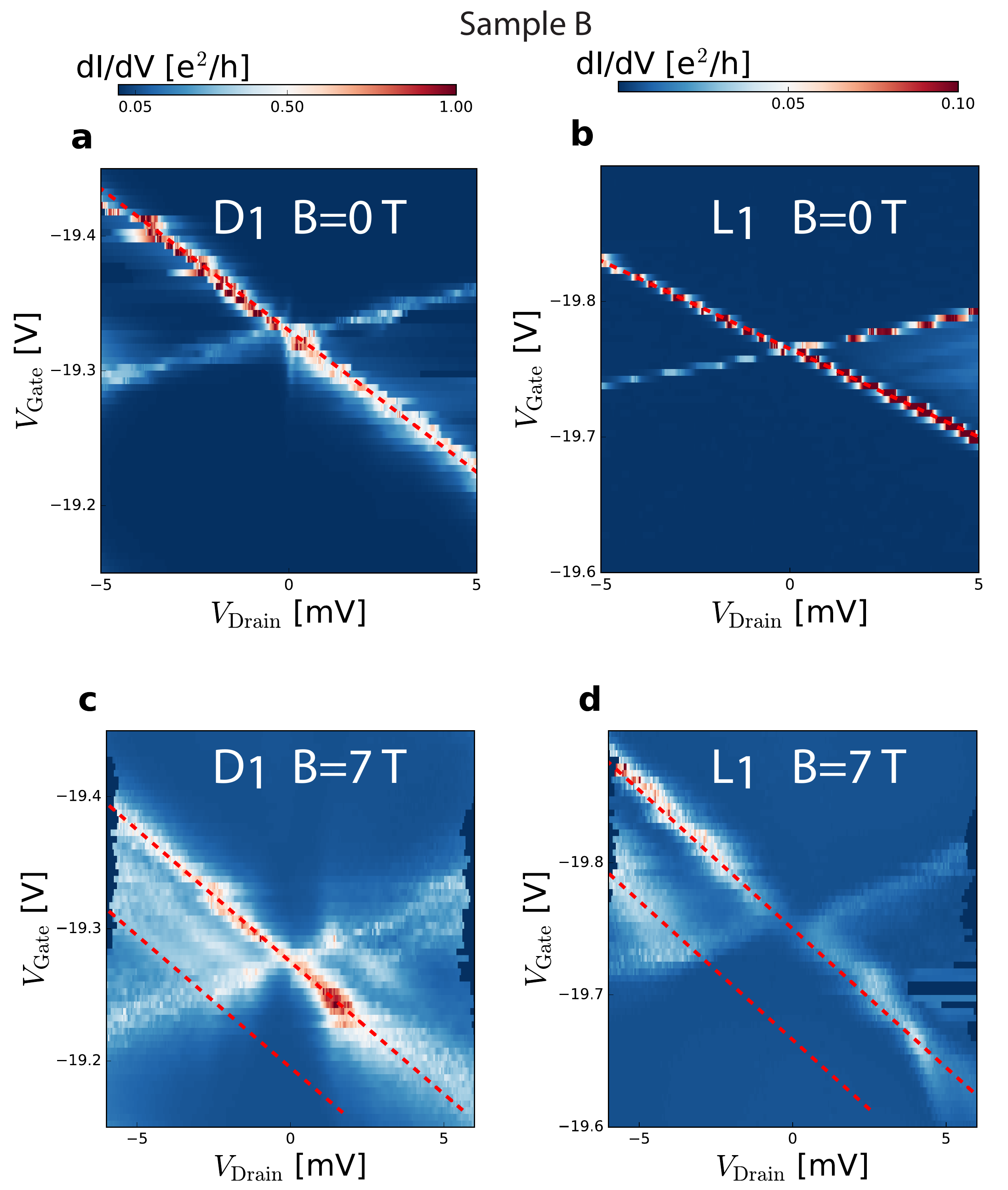}
		\caption{\label{FigS3} a) b) Zoom on the energy levels D1 and L1 of nanowire B measured at zero magnetic field. c) d) Zoom on the energy levels D1 and L1 of nanowire B measured at a magnetic field of B=7 T. A Zeeman splitting is observed from which a Lande factor $|$g$|\simeq 10$ is extracted.}
	\end{center}
\end{figure*}


\bibliography{Bibliography}